\let\includefigures=\iftrue
\let\useblackboard=\iftrue
\newfam\black
\input harvmac

\noblackbox

\includefigures
\message{If you do not have epsf.tex (to include figures),}
\message{change the option at the top of the tex file.}
\input epsf
\def\figin{\epsfcheck\figin}\def\figins{\epsfcheck\figins}
\def\epsfcheck{\ifx\epsfbox\UnDeFiNeD
\message{(NO epsf.tex, FIGURES WILL BE IGNORED)}
\gdef\figin##1{\vskip2in}\gdef\figins##1{\hskip.5in}
\else\message{(FIGURES WILL BE INCLUDED)}%
\gdef\figin##1{##1}\gdef\figins##1{##1}\fi}
\def\DefWarn#1{}
\def\figinsert{\goodbreak\midinsert}
\def\ifig#1#2#3{\DefWarn#1\xdef#1{fig.~\the\figno}
\writedef{#1\leftbracket fig.\noexpand~\the\figno}%
\figinsert\figin{\centerline{#3}}\medskip\centerline{\vbox{
\baselineskip12pt\advance\hsize by -1truein
\noindent\footnotefont{\bf Fig.~\the\figno:} #2}}
\bigskip\endinsert\global\advance\figno by1}
\else
\def\ifig#1#2#3{\xdef#1{fig.~\the\figno}
\writedef{#1\leftbracket fig.\noexpand~\the\figno}%
\global\advance\figno by1} \fi
%

\useblackboard
\message{If you do not have msbm (blackboard bold) fonts,}
\message{change the option at the top of the tex file.}
\font\blackboard=msbm10 scaled \magstep1 \font\blackboards=msbm7
\font\blackboardss=msbm5 \textfont\black=\blackboard
\scriptfont\black=\blackboards
\scriptscriptfont\black=\blackboardss

\else

\fi
%
\def\yboxit#1#2{\vbox{\hrule height #1 \hbox{\vrule width #1
\vbox{#2}\vrule width #1 }\hrule height #1 }}
\def\fillbox#1{\hbox to #1{\vbox to #1{\vfil}\hfil}}
\def\ybox{{\lower 1.3pt \yboxit{0.4pt}{\fillbox{8pt}}\hskip-0.2pt}}

\noblackbox
\includefigures
\message{If you do not have epsf.tex (to include figures),}
\message{change the option at the top of the tex file.}
\input epsf
\def\figin{\epsfcheck\figin}\def\figins{\epsfcheck\figins}
\def\epsfcheck{\ifx\epsfbox\UnDeFiNeD
\message{(NO epsf.tex, FIGURES WILL BE IGNORED)}
\gdef\figin##1{\vskip2in}\gdef\figins##1{\hskip.5in}
\else\message{(FIGURES WILL BE INCLUDED)}%
\gdef\figin##1{##1}\gdef\figins##1{##1}\fi}
\def\DefWarn#1{}
\def\figinsert{\goodbreak\midinsert}
\def\ifig#1#2#3{\DefWarn#1\xdef#1{fig. \the\figno}
\writedef{#1\leftbracket fig.\noexpand \the\figno}%
\figinsert\figin{\centerline{#3}}\medskip\centerline{\vbox{
\baselineskip12pt\advance\hsize by -1truein
\noindent\footnotefont{\bf Fig. \the\figno:} #2}}
\bigskip\endinsert\global\advance\figno by1}
\else
\def\ifig#1#2#3{\xdef#1{fig. \the\figno}
\writedef{#1\leftbracket fig.\noexpand \the\figno}%
\global\advance\figno by1} \fi
%

\useblackboard
\message{If you do not have msbm (blackboard bold) fonts,}
\message{change the option at the top of the tex file.}
\font\blackboard=msbm10 scaled \magstep1 \font\blackboards=msbm7
\font\blackboardss=msbm5 \textfont\black=\blackboard
\scriptfont\black=\blackboards
\scriptscriptfont\black=\blackboardss

\else

\fi
%
\def\yboxit#1#2{\vbox{\hrule height #1 \hbox{\vrule width #1
\vbox{#2}\vrule width #1 }\hrule height #1 }}
\def\fillbox#1{\hbox to #1{\vbox to #1{\vfil}\hfil}}
\def\ybox{{\lower 1.3pt \yboxit{0.4pt}{\fillbox{8pt}}\hskip-0.2pt}}

\def\IR{\relax{\rm I\kern-.18em R}}

\let\abs=|
%
%
\message{S-Tables Macro v1.0, ACS, TAMU (RANHELP@VENUS.TAMU.EDU)}
%
%
\newhelp\stablestylehelp{You must choose a style between 0 and 3.}%
\newhelp\stablelinehelp{You
should not use special hrules when stretching
a table.}%
\newhelp\stablesmultiplehelp{You have tried to place an S-Table
inside another
S-Table.  I would recommend not going on.}%
%
%
\newdimen\stablesthinline
\stablesthinline=0.4pt
\newdimen\stablesthickline
\stablesthickline=1pt
%
%
\newif\ifstablesborderthin
\stablesborderthinfalse
\newif\ifstablesinternalthin
\stablesinternalthintrue
\newif\ifstablesomit
\newif\ifstablemode
\newif\ifstablesright
\stablesrightfalse
%
%
\newdimen\stablesbaselineskip
\newdimen\stableslineskip
\newdimen\stableslineskiplimit
%
%
\newcount\stablesmode
\newcount\stableslines
\newcount\stablestemp
\stablestemp=3
\newcount\stablescount
\stablescount=0
\newcount\stableslinet
\stableslinet=0
%
%
%
\newcount\stablestyle
\stablestyle=0
%
%
\def\stablesleft{\quad\hfil}%
\def\stablesright{\hfil\quad}%
%
%
\catcode`\|=\active%
%
%
\newcount\stablestrutsize
\newbox\stablestrutbox
\setbox\stablestrutbox=\hbox{\vrule height10pt depth5pt width0pt}
\def\stablestrut{\relax\ifmmode%
                         \copy\stablestrutbox%
                       \else%
                         \unhcopy\stablestrutbox%
                       \fi}%
%
%
\newdimen\stablesborderwidth
\newdimen\stablesinternalwidth
\newdimen\stablesdummy
\newcount\stablesdummyc
\newif\ifstablesin
\stablesinfalse
%
%
\def\begintable{\stablestart%
  \stablemodetrue%
  \stablesadj%
  \halign%
  \stablesdef}%
\def\stablesadj{%
  \ifcase\stablestyle%
    \hbox to \hsize\bgroup\hss\vbox\bgroup%
  \or%
    \hbox to \hsize\bgroup\vbox\bgroup%
  \or%
    \hbox to \hsize\bgroup\hss\vbox\bgroup%
  \or%
    \hbox\bgroup\vbox\bgroup%
  \else%
    \errhelp=\stablestylehelp%
    \errmessage{Invalid style selected, using default}%
    \hbox to \hsize\bgroup\hss\vbox\bgroup%
  \fi}%
\def\stablesend{\egroup%
  \ifcase\stablestyle%
    \hss\egroup%
  \or%
    \hss\egroup%
  \or%
    \egroup%
  \or%
    \egroup%
  \else%
    \hss\egroup%
  \fi}%
\def\stablestart{%
  \ifstablesin%
    \errhelp=\stablesmultiplehelp%
    \errmessage{An S-Table cannot be placed within an S-Table!}%
  \fi
  \global\stablesintrue%
  \global\advance\stablescount by 1%
  \message{<S-Tables Generating Table \number\stablescount}%
  \begingroup%
  \stablestrutsize=\ht\stablestrutbox%
  \advance\stablestrutsize by \dp\stablestrutbox%
  \ifstablesborderthin%
    \stablesborderwidth=\stablesthinline%
  \else%
    \stablesborderwidth=\stablesthickline%
  \fi%
  \ifstablesinternalthin%
    \stablesinternalwidth=\stablesthinline%
  \else%
    \stablesinternalwidth=\stablesthickline%
  \fi%
  \tabskip=0pt%
  \stablesbaselineskip=\baselineskip%
  \stableslineskip=\lineskip%
  \stableslineskiplimit=\lineskiplimit%
  \offinterlineskip%
  \def\borderrule{\vrule width \stablesborderwidth}%
  \def\internalrule{\vrule width \stablesinternalwidth}%
  \def\thinline{\noalign{\hrule height \stablesthinline}}%
  \def\thickline{\noalign{\hrule height \stablesthickline}}%
  \def\trule{\omit\leaders\hrule height \stablesthinline\hfill}%
  \def\ttrule{\omit\leaders\hrule height \stablesthickline\hfill}%
  \def\tttrule##1{\omit\leaders\hrule height ##1\hfill}%
  \def\stablesel{&\omit\global\stablesmode=0%
    \global\advance\stableslines by 1\borderrule\hfil\cr}%
  \def\el{\stablesel&}%
  \def\elt{\stablesel\thinline&}%
  \def\eltt{\stablesel\thickline&}%
  \def\elttt##1{\stablesel\noalign{\hrule height ##1}&}%
  \def\elspec{&\omit\hfil\borderrule\cr\omit\borderrule&%
              \ifstablemode%
              \else%
                \errhelp=\stablelinehelp%
                \errmessage{Special ruling will not display properly}%
              \fi}%
  \def\stmultispan##1{\mscount=##1 \loop\ifnum\mscount>3
\stspan\repeat}%
  \def\stspan{\span\omit \advance\mscount by -1}%
  \def\multicolumn##1{\omit\multiply\stablestemp by ##1%
     \stmultispan{\stablestemp}%
     \advance\stablesmode by ##1%
     \advance\stablesmode by -1%
     \stablestemp=3}%
  \def\multirow##1{\stablesdummyc=##1\parindent=0pt\setbox0\hbox\bgroup%
    \aftergroup\emultirow\let\temp=}
  \def\emultirow{\setbox1\vbox to\stablesdummyc\stablestrutsize%
    {\hsize\wd0\vfil\box0\vfil}%
    \ht1=\ht\stablestrutbox%
    \dp1=\dp\stablestrutbox%
    \box1}%

\def\stpar##1{\vtop\bgroup\hsize ##1%
     \baselineskip=\stablesbaselineskip%
     \lineskip=\stableslineskip%

\lineskiplimit=\stableslineskiplimit\bgroup\aftergroup\estpar\let\temp=}%
  \def\estpar{\vskip 6pt\egroup}%
  \def\stparrow##1##2{\stablesdummy=##2%
     \setbox0=\vtop to ##1\stablestrutsize\bgroup%
     \hsize\stablesdummy%
     \baselineskip=\stablesbaselineskip%
     \lineskip=\stableslineskip%
     \lineskiplimit=\stableslineskiplimit%
     \bgroup\vfil\aftergroup\estparrow%
     \let\temp=}%
  \def\estparrow{\vfil\egroup%
     \ht0=\ht\stablestrutbox%
     \dp0=\dp\stablestrutbox%
     \wd0=\stablesdummy%
     \box0}%
  \def|{\global\advance\stablesmode by 1&&&}%
  \def\|{\global\advance\stablesmode by 1&\omit\vrule width 0pt%
         \hfil&&}%
  \def\vt{\global\advance\stablesmode by 1&\omit\vrule width
\stablesthinline%
          \hfil&&}%
  \def\vtt{\global\advance\stablesmode by 1&\omit\vrule width
\stablesthickline%
          \hfil&&}%
  \def\vttt##1{\global\advance\stablesmode by 1&\omit\vrule width ##1%
          \hfil&&}%
  \def\vtr{\global\advance\stablesmode by 1&\omit\hfil\vrule width%
           \stablesthinline&&}%
  \def\vttr{\global\advance\stablesmode by 1&\omit\hfil\vrule width%
            \stablesthickline&&}%
  \def\vtttr##1{\global\advance\stablesmode by 1&\omit\hfil\vrule
width ##1&&}%
  \stableslines=0%
  \stablesomitfalse}
\def\stablesdef{\bgroup\stablestrut\borderrule##\tabskip=0pt plus 1fil%
  &\stablesleft##\stablesright%
  &##\ifstablesright\hfill\fi\internalrule\ifstablesright\else\hfill\fi%
  \tabskip 0pt&&##\hfil\tabskip=0pt plus 1fil%
  &\stablesleft##\stablesright%
  &##\ifstablesright\hfill\fi\internalrule\ifstablesright\else\hfill\fi%
  \tabskip=0pt\cr%
  \ifstablesborderthin%
    \thinline%
  \else%
    \thickline%
  \fi&%
}%
\def\endtable{\advance\stableslines by 1\advance\stablesmode by 1%
   \message{- Rows: \number\stableslines, Columns:
\number\stablesmode>}%
   \stablesel%
   \ifstablesborderthin%
     \thinline%
   \else%
     \thickline%
   \fi%
   \egroup\stablesend%
\endgroup%
\global\stablesinfalse}
%

\def\hh{{1\over 2}}

\def\ll{_}
\def\uu{^}
\def\pp{\partial}

\def\exp#1{{\rm exp}\{#1\}}
\def\d{\delta}
\def\m{\mu}

\def\dag{{}^\dagger{}}

\def\m{\mu}
\def\n{\nu}
\def\s{\sigma}
\def\t{\tau}
\def\G{\Gamma}
\def\g{\gamma}
\def\a{\alpha}
\def\r{\rho}

\def\o{\omega}
\def\e{\epsilon}

\def\sqd{^2}
\def\zb{{\bar{z}}}

\def\hh{{1\over 2}}

\def\ee#1{\eqn\placeholder {\eqalign{#1}}}

\def\th{\theta}
\def\t{\tau}

\def\ww{\wedge}

\def\b{\beta}

\def\llsk{\hskip .5in}

\def\st{{}^*}

\def\lsq{\left [}
\def\rsq{\right ]}

\def\pr{^\prime {}}

\def\apr{{\alpha^\prime} {}}

\def\ww{\wedge}
\def\IZ{\relax\ifmmode\mathchoice
{\hbox{\cmss Z\kern-.4em Z}}{\hbox{\cmss Z\kern-.4em Z}}
{\lower.9pt\hbox{\cmsss Z\kern-.4em Z}} {\lower1.2pt\hbox{\cmsss
Z\kern-.4em Z}}\else{\cmss Z\kern-.4em Z}\fi} \font\cmss=cmss10
\font\cmsss=cmss10 at 7pt
\def\inbar{\,\vrule height1.5ex width.4pt depth0pt}
\def\IC{{\relax\hbox{$\inbar\kern-.3em{\rm C}$}}}
\def\IQ{{\relax\hbox{$\inbar\kern-.3em{\rm Q}$}}}
\def\IP{\relax{\rm I\kern-.18em P}}

\def\k#1{{k_{#1}}}

\def\Psb{\bar{\Psi}}
\def\k{\kappa}
\def\l{{\tilde{\lambda}}}

\def\ket#1{|{#1}\rangle}

\def\lrd{\left (}
\def\rrd{\right )}

\def\alt{\tilde{\alpha}}

\def\htt{\tilde{h}}
\def\ot{\otimes}

\def\hop#1{{\rm HO}^{+{#1}}{}}
\def\ho#1{{\rm HO}^{{#1}}{}}
\def\hod#1{{\rm HO}^{{#1}}{}^/ {}}

\def\psb{\bar{\psi}}
\def\mc{M$^{\underline{\rm c}}$}

\def\ups{\Upsilon}

\def\ww{\wedge}

\def\ff{{1\over 4}}

\def\o{\omega}

\def\ct{\tilde{c}}

\def\thb{\bar{\theta}}

\def\pr{^\prime{}}

\def\at{\tilde{\alpha}}
\def\gt#1#2{\tilde{G}^{(#1)}_{#2}}
\def\pst{\tilde{\psi}}

\def\cl{{\cal L}}

\def\at{{\tilde{\alpha}}}
\def\phn{\Phi_0}
\def\cht{{\tilde{\chi}}}
\def\Gt{\tilde{\Gamma}}
\def\ald{{\dot{\alpha}}}
\def\bed{\dot{\beta}}

\def\gt{\tilde{\gamma}}
\def\pd{{\dot{p}}}
\def\qd{{\dot{q}}}
\def\ct{{\cal T}}
\def\cx{{\bf X}}
\def\sgn{{\rm sign}}
\def\dt#1#2{{{\partial T^{#1}}\over{\partial x^{#2}}}}
\def\supercrit{\ChamseddineFG, \MyersFV, \deAlwisPR}
\def\Psh{{\hat{\Psi}}}
\def\psh{{\hat{\psi}}}
\def\thh{{\hat{\theta}}}
\def\upsh{{\hat{\Upsilon}}}
\def\upsb{\bar{\Upsilon}}
\def\dnb{\overline{\rm D9}}

\def\ad#1{\overline{\rm D{#1}}}

\def\cv{{\cal V}}
\def\cct{{\tilde{c}}}

\def\dbra#1{\left \langle \left \langle {#1} \right \abs \right .}
\def\ket#1{\left\abs {#1} \right \rangle}
\def\spherecor#1#2#3{\left\langle ~ {#1}(z_1)~{#2}(z_2)~{#3}(z_3)
~\right \rangle_{S^2}}

\lref\AdamsSV{
A.~Adams, J.~Polchinski and E.~Silverstein,
``Don't panic! Closed string tachyons in ALE space-times,''
JHEP {\bf 0110}, 029 (2001)
[arXiv:hep-th/0108075].
}

\lref\PolchinskiDF{
J.~Polchinski and E.~Witten,
``Evidence for Heterotic - Type I String Duality,''
Nucl.\ Phys.\ B {\bf 460}, 525 (1996)
[arXiv:hep-th/9510169].
}

\lref\BlauPM{
M.~Blau,
``The Mathai-Quillen formalism and topological field theory,''
J.\ Geom.\ Phys.\  {\bf 11}, 95 (1993)
[arXiv:hep-th/9203026].
}

\lref\fzzunpublished{
V. Fateev, A. Zamolodchikov and Al. Zamolodchikov, unpublished.}

\lref\TeschnerFT{
J.~Teschner,
``On structure constants and fusion rules in the SL(2,C)/SU(2) WZNW  model,''
Nucl.\ Phys.\ B {\bf 546}, 390 (1999)
[arXiv:hep-th/9712256].
}

\lref\TeschnerFV{
J.~Teschner,
``The mini-superspace limit of the SL(2,C)/SU(2) WZNW model,''
Nucl.\ Phys.\ B {\bf 546}, 369 (1999)
[arXiv:hep-th/9712258].
}

\lref\TeschnerUG{
J.~Teschner,
``Operator product expansion and factorization in the H-3+ WZNW model,''
Nucl.\ Phys.\ B {\bf 571}, 555 (2000)
[arXiv:hep-th/9906215].
}
\lref\KazakovPM{
V.~Kazakov, I.~K.~Kostov and D.~Kutasov,
``A matrix model for the two-dimensional black hole,''
Nucl.\ Phys.\ B {\bf 622}, 141 (2002)
[arXiv:hep-th/0101011].
}

\lref\MooreGA{
G.~W.~Moore,
``Gravitational phase transitions and the Sine-Gordon model,''
arXiv:hep-th/9203061.
}

\lref\BergNE{
M.~Berg, C.~DeWitt-Morette, S.~Gwo and E.~Kramer,
``The Pin groups in physics: C, P, and T,''
Rev.\ Math.\ Phys.\  {\bf 13}, 953 (2001)
[arXiv:math-ph/0012006].
}

\lref\ChamseddineFG{
A.~H.~Chamseddine,
``A Study of noncritical strings in arbitrary dimensions,''
Nucl.\ Phys.\ B {\bf 368}, 98 (1992).
}

\lref\SuyamaAS{
T.~Suyama,
``On decay of bulk tachyons,''
arXiv:hep-th/0308030.
}

\lref\HorowitzGN{
G.~T.~Horowitz and L.~Susskind,
``Bosonic M theory,''
J.\ Math.\ Phys.\  {\bf 42}, 3152 (2001)
[arXiv:hep-th/0012037].
}

\lref\SeibergRS{
N.~Seiberg and E.~Witten,
``Electric - magnetic duality, monopole condensation, and confinement in N=2
supersymmetric Yang-Mills theory,''
Nucl.\ Phys.\ B {\bf 426}, 19 (1994)
[Erratum-ibid.\ B {\bf 430}, 485 (1994)]
[arXiv:hep-th/9407087].
}

\lref\BerkovitsXQ{
N.~Berkovits and C.~Vafa,
``On the Uniqueness of string theory,''
Mod.\ Phys.\ Lett.\ A {\bf 9}, 653 (1994)
[arXiv:hep-th/9310170].
}

\lref\SilversteinXN{
E.~Silverstein,
``(A)dS backgrounds from asymmetric orientifolds,''
arXiv:hep-th/0106209.
}

\lref\MaloneyRR{
A.~Maloney, E.~Silverstein and A.~Strominger,
``De Sitter space in noncritical string theory,''
arXiv:hep-th/0205316.
}

\lref\StromingerFN{
A.~Strominger and T.~Takayanagi,
``Correlators in timelike bulk Liouville theory,''
Adv.\ Theor.\ Math.\ Phys.\  {\bf 7}, 369 (2003)
[arXiv:hep-th/0303221].
}

\lref\PolchinskiRQ{
J.~Polchinski,
``String Theory. Vol. 1: An Introduction To The Bosonic String.''
}

\lref\PolchinskiRR{
J.~Polchinski,
``String Theory. Vol. 2: Superstring Theory And Beyond.''
}

\lref\MyersFV{
R.~C.~Myers,
``New Dimensions For Old Strings,''
Phys.\ Lett.\ B {\bf 199}, 371 (1987).
}

\lref\deAlwisPR{
S.~P.~de Alwis, J.~Polchinski and R.~Schimmrigk,
``Heterotic Strings With Tree Level Cosmological Constant,''
Phys.\ Lett.\ B {\bf 218}, 449 (1989).
}

\lref\SchwarzSF{
J.~H.~Schwarz and E.~Witten,
``Anomaly analysis of brane-antibrane systems,''
JHEP {\bf 0103}, 032 (2001)
[arXiv:hep-th/0103099].
}

\lref\FlournoyVN{
A.~Flournoy, B.~Wecht and B.~Williams,
``Constructing nongeometric vacua in string theory,''
arXiv:hep-th/0404217.
}

\lref\SenVD{
A.~Sen,
``F-theory and Orientifolds,''
Nucl.\ Phys.\ B {\bf 475}, 562 (1996)
[arXiv:hep-th/9605150].
}

\lref\HellermanBU{
S.~Hellerman, S.~Kachru, A.~E.~Lawrence and J.~McGreevy,
``Linear sigma models for open strings,''
JHEP {\bf 0207}, 002 (2002)
[arXiv:hep-th/0109069].
}

\lref\HellermanCT{
S.~Hellerman and J.~McGreevy,
``Linear sigma model toolshed for D-brane physics,''
JHEP {\bf 0110}, 002 (2001)
[arXiv:hep-th/0104100].
}

\lref\FischlerCI{
W.~Fischler and L.~Susskind,
``Dilaton Tadpoles, String Condensates And Scale Invariance,''
Phys.\ Lett.\ B {\bf 171}, 383 (1986).
}

\lref\FischlerTB{
W.~Fischler and L.~Susskind,
``Dilaton Tadpoles, String Condensates And Scale Invariance. 2,''
Phys.\ Lett.\ B {\bf 173}, 262 (1986).
}

\lref\FischlerGZ{
W.~Fischler, I.~R.~Klebanov and L.~Susskind,
``String Loop Divergences And Effective Lagrangians,''
Nucl.\ Phys.\ B {\bf 306}, 271 (1988).
}

\lref\Milnor{
J.~W.~Milnor and J.~D.~Stasheff, ``Characteristic
Classes''.}

\lref\Anderson{D.~W.~Anderson, E.~H.~Brown and F.~P.~Peterson,
``The Structure of the Spin Cobordism Ring,''
Ann.\ Math. {\bf 86}, 271 (1967).}

\lref\KachruED{
S.~Kachru, J.~Kumar and E.~Silverstein,
``Orientifolds, RG flows, and closed string tachyons,''
Class.\ Quant.\ Grav.\  {\bf 17}, 1139 (2000)
[arXiv:hep-th/9907038].
}

\lref\Hellermantoappear{S. Hellerman and J. \mc Greevy, to appear.}

\Title{\vbox{\baselineskip12pt\hbox{hep-th/0405041}}}
{\vbox{
\centerline{On the Landscape}
\bigskip
\centerline{of}
\bigskip
\centerline{Superstring Theory in D $>$ 10}
}}
\bigskip
\bigskip
\centerline{Simeon Hellerman$^{1}$}
\bigskip
\centerline{$^{1}${\it School of Natural Sciences,
Institue for Advanced Study,
Princeton, NJ 08540}}
\smallskip
\bigskip
\bigskip
\noindent

We study a family of unstable heterotic string theories
in more than ten dimensions which are
connected via tachyon condensation to the
ten-dimensional supersymmetric vacuum of
heterotic string theory with gauge group
$SO(32)$.  Calculating the spectrum
of these theories, we find evidence for an S-duality
which relates type I string theory in ten dimensions with 
$n$ additional ninebrane-antininebrane pairs
to heterotic string
theory in $10+n$ dimensions with gauge group $SO(32+n)$.
The Kaluza-Klein modes of the supercritical dimensions
are dual to non-singlet bound states of open strings.

\bigskip
\Date{May 5, 2004}

\newsec{Introduction}

Superstring theories formulated in dimensions greater than ten have played
little role in our understanding of the nonperturbative structure
of string theory.  In particular, the dualities among
string theories have
so far connected only a closed set containing the
theories in which the superstring propagates in the critical number
of spacetime dimensions.

At the perturbative level
noncritical string theories are equally consistent
\ChamseddineFG, \MyersFV, \deAlwisPR.
In the background of a dilaton with
gradient of appropriate magnitude,
strings in supercritical dimensions
propagate consistently with the rules of quantum mechanics.

Unlike their critical cousins, supercritical superstring
theories lack linearly
realized spacetime supersymmetry and as a result
they are vulnerable to
tadpoles,
mass renormalizations, and other perturbative instabilities, but such problems
do not fundamentally destroy the consistency of
string perturbation theory \FischlerCI, \FischlerGZ, \FischlerTB.
The supercritical theories are just as
sensible in perturbation theory as the critical ones, and yet
we lack a comparable understanding of their nonperturbative behavior.

In this paper we will study two families of
supercritical heterotic strings with orthogonal gauge
group, which we will call
$\ho +$ and $\ho {+/}$.
Both are perturbatively unstable against
decay to lower dimensions.  Tachyon condensation
in the $\ho +$ theory will generically lead to 
a ten-dimensional supersymmetric string theory.
The $\ho{+/}$ theory is more unstable.  It can also decay
to ten dimensions, but $\ho{+/}$ theory on a smooth
space cannot decay to a supersymmetric theory, or
even a stable one.  On a space with certain orbifold
singularities, $\ho {+/}$ can decay to the supersymmetric
ten-dimensional HO background; the stabilization
relies on the boundary condition of the $\ho {+/}$
bulk tachyons at
the orbifold singularity.

Examining the chiral fermion spectrum at the singularity,
we will find that it matches the chiral fermion
spectrum of type I string theory in ten dimensions, with
additional D9-$\ad 9$ pairs added.  We will be led to
propose that a suitable compactification of
the $\ho{+/}$ theory, with a single orbifold singularity
of the correct type, plays the role of a strong
coupling limit of the unstable type I theory.
If correct, this duality provides
the first link between the landscape of
critical string backgrounds and the supercritical world which
lies beyond.  

\ifig\susyweb{There are six simple limits of string theory
with ten- (or higher-) dimensional super-Poincar\'e symmetry,
connected by known dualities and deformations.
}
{\epsfxsize3.0in\epsfbox{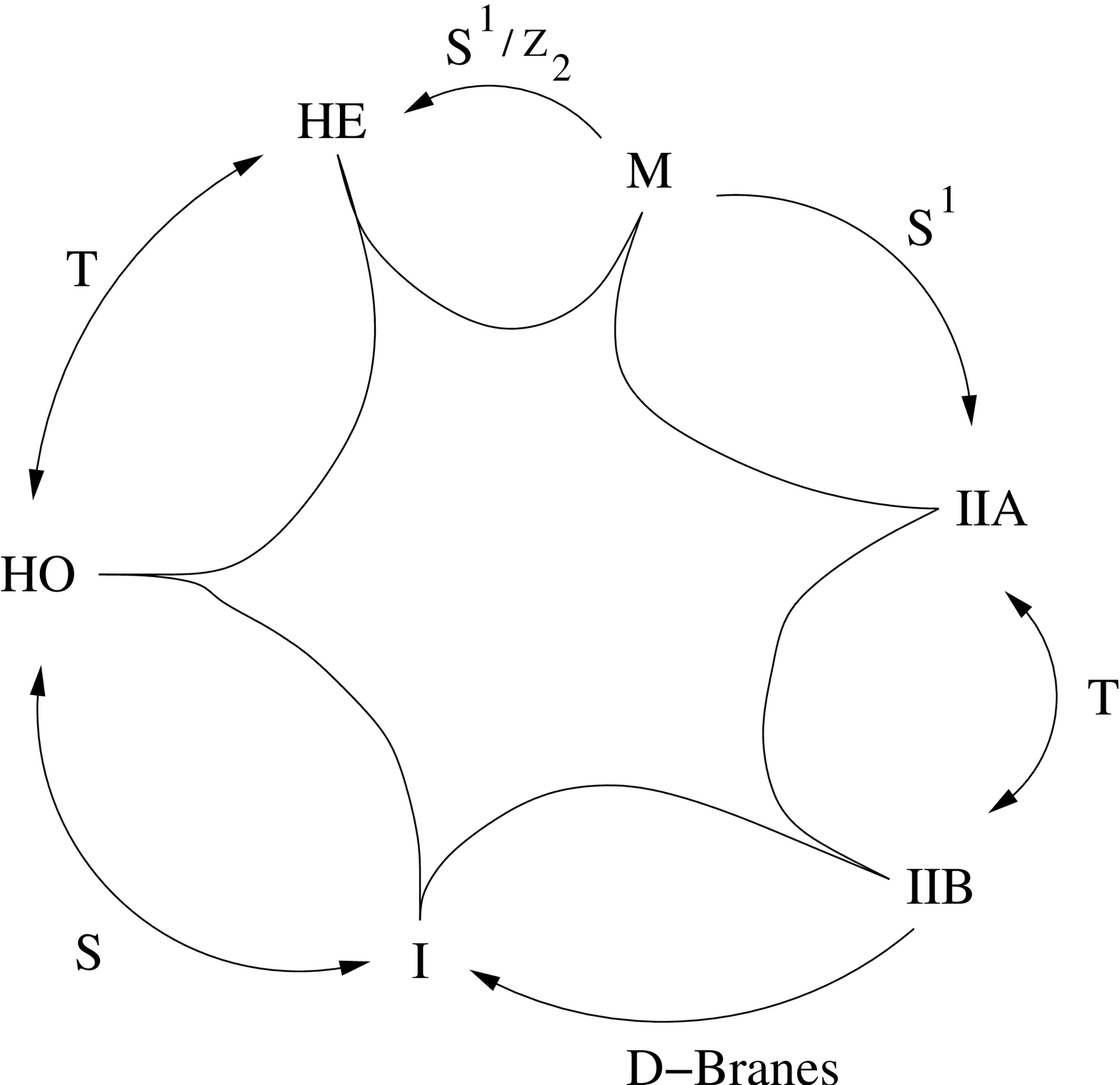}}

\ifig\nonsusyweb{
Nine-dimensionally Poincar\'e-invariant configurations
of string theory which lie above
the moduli space.  
We will fill in
the empty spot on the phase diagram with a
supercritical heterotic string theory,
which is exchanged by S-duality with the type
I+n~D9 + n$~\overline{{\rm D9}}$ vacuum.
}
{\epsfxsize3.0in\epsfbox{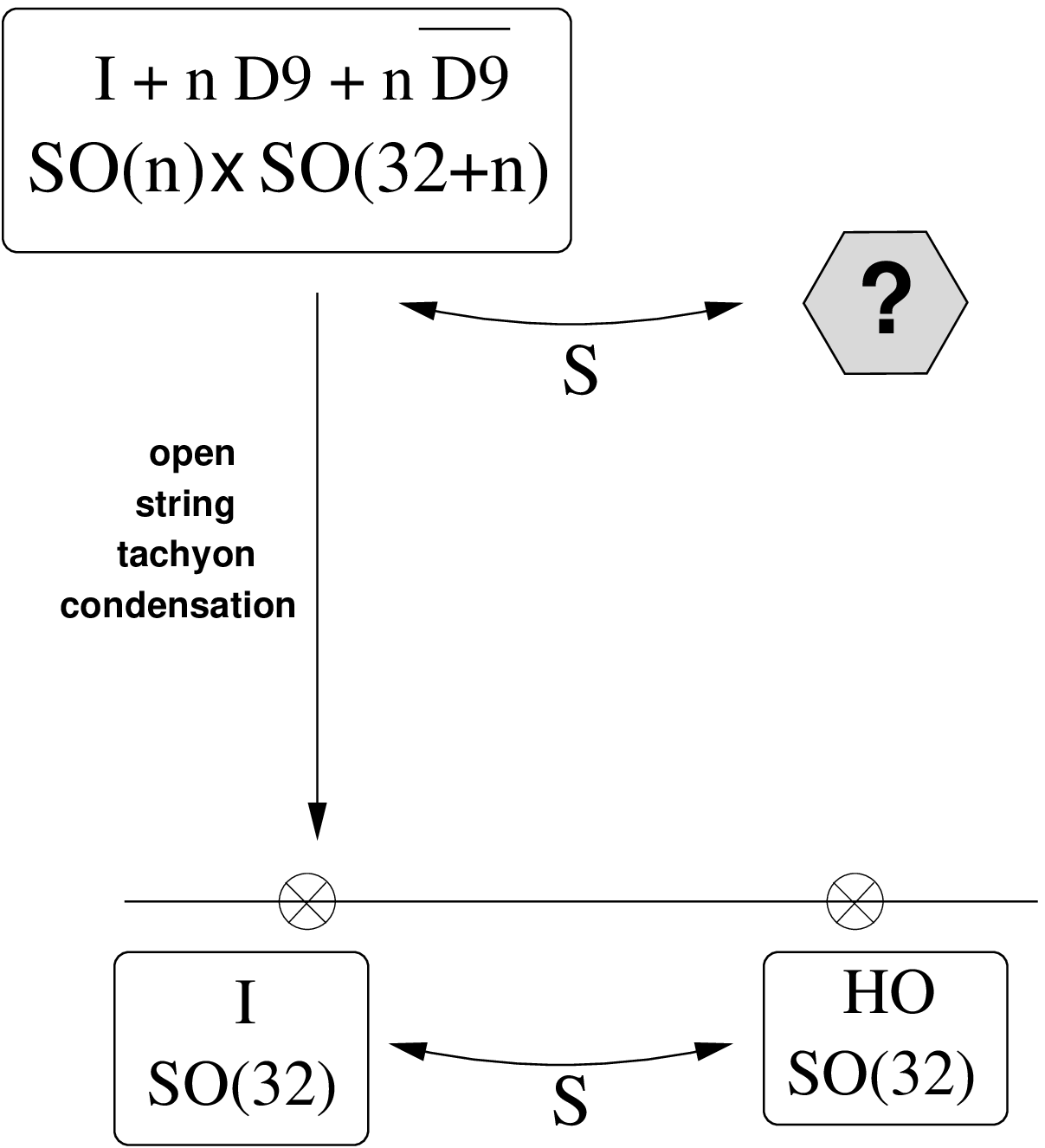}}

The organization of the paper is as follows.  In section 2, we
introduce the reader to the basic ideas of
superstring theory in more than ten dimensions.  In section
3, we present two families
of supercritical heterotic string theories,
$\ho +$ and $\ho {+/}$, which are connected 
to the ten-dimensional $SO(32)$ heterotic
superstring theory by tachyon condensation.
(Attempts to understand the endpoint of closed string
tachyon condensation in other string
theories include \AdamsSV, \HorowitzGN, \KachruED,
\SuyamaAS.)
In sections 3 and
4, 
we explore local and global aspects of tachyon condensation
in $\ho +$.
In section 5, we consider $\ho {+/}$ on
orbifolds, with particular attention to the
spectrum of chiral fermions at the fixed locus
and its correspondence with the fermion spectrum in
an unstable type I background.
In section 6, we sharpen our S-duality conjecture in
certain cases
by studying $\ho{+/}$ on compact toroidal orbifolds
whose ten-dimensional spectrum of chiral fermions matches 
that of the type I theory with $n=1,2$ ninebrane-antininebrane
pairs.

\newsec{The basics of supercritical string theory}

The basic ideas which allow a consistent interpretation
of the supercritical string appear in \ChamseddineFG ~; in
this section we review them.

We give particular emphasis to the point that
the supercritical string reproduces completely
conventional spacetime physics.  Despite the
presence of a dilaton gradient with string-scale
magnitude, spacetime processes are faithfully
described by a local effective action (\ChamseddineFG,
\SilversteinXN, \MaloneyRR, \deAlwisPR, \MyersFV).
 
\subsec{A simple example -- maximally Lorentz-invariant
type II strings}

The physical state conditions and dispersion
relations for strings in linear dilaton backgrounds
have been analyzed in detail (\ChamseddineFG, \deAlwisPR, \MyersFV).
For an illustrative example
we now analyze the NS sector of the supercritical type II string.
The type II string with
Lorentz-invariant chiral GSO projection can exist only in dimensions
equal to $10$ mod $8$ (\MyersFV, \ChamseddineFG).

The Virasoro generators are

\ee{
L\uu{mat.}\ll n - &A\uu{mat.} \d\ll{0,n} =
\cr
\hh \sum\ll{m} :~\a\uu\m\ll{n-m} \a\ll{\m m} ~:
+ \ff \sum\ll{r} (2r-m) &:~\psi\uu\m\ll{n-r} \psi\ll{\m r} ~:
+ i \left ( {\apr \over 2} \right )\uu\hh (n+1)
V\ll\m \a\uu\m\ll n
\cr
&
\cr
G\uu{mat.}\ll r = \sum\ll n \a\uu\m\ll n \psi\ll{\m~r-n}
+ &i \lrd {\apr\over 2} \rrd\uu\hh (2r+1)V\ll\m \psi\uu\m\ll r
\cr
&
\cr
\tilde{L}\uu{mat.}\ll n  - &\tilde{A}\uu{mat.} \d\ll{0,n} =
\cr
\hh \sum\ll{m} :~\tilde{\a}\uu\m\ll{n-m} \tilde{\a}\ll{\m m} ~:
+
 \ff \sum\ll{r} (2r-m) &:~\tilde{\psi}\uu\m 
\ll{n-r} \tilde{\psi}\ll {\m~r}  ~:
+ i \left ( {\apr \over 2} \right )\uu\hh (n+1)
V\ll\m \tilde{\a}\uu\m\ll n
\cr
&
\cr
\tilde{G}\uu{mat.}\ll r = \sum\ll n \alt\uu\m\ll n \pst\ll{\m~r-n}
+ &i \lrd {\apr\over 2} \rrd\uu\hh (2r+1)V\ll\m \pst\uu\m\ll r
}

As always, superghosts contribute $-1$ to the fermion number
of the ground state; so the lowest mode surviving the
GSO projection in the NS/NS sector is
\ee{
\left \abs  \Psi (e,k) \right \rangle
\equiv e\ll{\m\n} \pst\ll{-\hh} \uu\m \psi\ll{-\hh}\uu\n \left \abs  k;0\right \rangle
\ll{{\rm NS/NS}}
}

We examine the dispersion relations, physical state
conditions, and gauge equivalences of this level.
Let us consider the simplest case, where $e\ll{\m\n}$
is antisymmetric in $\m$ and $\n$.
The gauge invariance
\ee{
\left \abs  \Psi \right \rangle \to \left \abs  \Psi \right \rangle + 
\tilde{G}\ll{-\hh}\uu{mat.} \left \abs  {\cal G}
\uu{(-\hh,0)} \right \rangle + G\uu{mat.}\ll{-\hh} \left \abs  {\cal G}\uu{(0,-\hh)} \right \rangle
}
acts on the state as
\ee{
e\ll{\m\n} \to e\ll{\m\n} + k \ll\m \Lambda\ll\n - k\ll\n 
\Lambda\ll\m,
}
which is the Fourier-space version of the usual 
gauge transformation
\ee{
B\ll{\m\n}(x) \to B\ll{\m\n}(x)
 + \Lambda\ll{\n,\m}(x) - \Lambda\ll{\m,\n}
(x).
}
We have used the closed-string identification 
\ee{
\a\ll 0\uu\m = \lrd {\apr\over 2} \rrd\uu \hh k\uu\m.
}
The gauge parameters $\left \abs  {\cal G}
\uu{(-\hh,0)} \right \rangle$ and $\left \abs  {\cal G}\uu{(0,-\hh)} \right \rangle$ can be written as
\ee{
\left \abs  {\cal G}
\uu{(-\hh,0)} \right \rangle = 
 (l\ll\m + \Lambda\ll\m)\psi\uu\m\ll{-\hh} \left \abs
k,0\right \rangle\ll{\rm NS/NS}
}
and
\ee{
 \left \abs  {\cal G}\uu{(0,-\hh)} \right \rangle = 
(l\ll\m - \Lambda\ll\m)\pst\uu\m\ll{-\hh}
 \left \abs
k,0\right \rangle\ll{\rm NS/NS},
}
where $l\uu\m$ is the vector field parametrizing an
infinitesimal coordinate transformation.

The gauge transformation of the string state
$\left \abs \Psi(e,k) \right \rangle$ has no
dependence on $V\ll\m$.  This gives us information about
the dilaton dependence of its normalization.
The state must be interpreted as a linear
fluctuation
of the $B\ll{\m\n}$ field itself, rather than the
field $\exp{-\Phi\ll 0} B\ll{\m\n}$ with canonical kinetic term.
If the string state represented the canonical field,
the gauge transformation law would be different, with
a nontrivial dependence on the background dilaton
gradient $V\ll\m$ coming from the action of $l\ll\m$ on
the background value $\Phi\ll 0$ of the dilaton.

The transversality conditions coming from
the $\tilde{G}\uu{mat.}\ll{\hh}, G\uu{mat.}\ll{\hh}$ physical state conditions,
do depend on the background value of the dilaton, and
they amount to
\ee{
(k + 2 i V)\uu\m ~ e\ll{\m\n} = 
k\pr \uu\m ~ e\ll{\m\n} = 0,
} 
where we define $k\pr \equiv k + 2 i  V$.
The dispersion relation, which comes from the conditions
\ee{
L\ll 0\uu{mat.} = \left \{ \matrix { \hh, & {\rm (NS)} \cr
                                          &  \cr
				     {5\over 8}, &  
                                     {\rm (R)} } \right \}
\cr
\tilde{L}\ll 0\uu{mat.} = \left \{\matrix{\hh,&(\widetilde{{\rm NS}})
                                             \cr
                                          &  \cr
				     {5\over 8}, &  
                                    (\widetilde{{\rm R} })} \right \}
}
and the fact that
\ee{
A\uu{mat.}  =
                           \left \{ \matrix { 0, & {\rm (NS)} \cr
                                          &  \cr
				     {D\over{16}}, &  
                                     {\rm (R)} } \right \}
\cr
\tilde{A}\uu{mat.} = \left \{\matrix{0,&(\widetilde{{\rm NS}})
                                             \cr
                                          &  \cr
				     {D\over {16}}, &  
                                    (\widetilde{{\rm R} })} \right \}
}
says that
\ee{
k\sqd + 2i V\cdot k = k\cdot k\pr = 0
}
for this state.  At first it appears puzzling that the 
dispersion relation is not Lorentz invariant, until we recall
(\ChamseddineFG, \deAlwisPR, \MyersFV)
that the $B$ field appears in the
string action as
\ee{
{\cal L}\ll B = {1\over{2\k\ll D\sqd}} \sqrt{\abs  G\abs  } 
~\exp{-2\Phi} \lsq -{1\over{12}} 
H\ll{\m\n\s} H\uu{\m\n\s}  \rsq  ,
}
and that the dilaton $\Phi$ has a background value
\ee{
\Phi\ll 0  = {\rm const.} + V\ll\m X\uu\m
}

If we linearize the equations of motion about this background,
we recover the B-field EOM
\ee{
(\pp\uu\s + 2 V\uu\s ) (\pp\ll \s B\ll{\m\n} + ({\rm cyclic})) = 0.
}
Together with the real-space transversality condition
\ee{
(\pp\uu\m + 2 V\uu\m) B\ll{\m\n} = 0
}
which we derived above from the $L\ll 1$ constraint, 
the EOM for $B\ll {\m\n}$
is the same as the dispersion relation we obtained from
the $L\ll 0$ (and $\tilde{L}\ll 0$) condition.
A similar analysis goes through
for the other NS/NS ground states \ChamseddineFG.

\newsec{The landscape of supercritical heterotic string theory}

\subsec{Two heterotic string theories in eleven dimensions}

There are two heterotic string theories with manifest
eleven-dimensional Poincar\'e invariance which will be
relevant for us.

The worldsheet degrees of freedom are
eleven free embedding coordinates $X\uu\m$ and their
superpartners $\psi\uu\m\ll +$ under the right-moving
supercharge $Q\ll +$, as well as thirty two left-moving
fermions
$\l\uu a\ll -$, with $a = 1,\cdots,32$ and a thirty-third
left-moving fermion $\cht$.

We call this theory $\ho {+(1)}$.  The plus stands for the fact that
this theory is like the critical HO theory, only with
more degrees of freedom on the worldsheet.  The $1$ stands
for the fact that there is a single extra dimension.

The $\ho{+(1)}$ theory 
has symmetry group ${\rm SO} (32) \times {\rm O}(10,1) $.
(The second
factor is spontaneously broken by the dilaton gradient
but by 
nothing else.)  Unlike the ten-dimensional HO theory,
the $\ho {+(1)}$ theory is invariant under
orientation-reversing orthogonal transformations of space.

If there are $n$ extra dimensions, we will
refer to the theory as type $\ho{+(n)}$.
When we add $n$ extra embedding coordinates,
they will have $n$ right-moving fermionic superpartners.
The number of extra left-moving fermions $\cht\uu A$
must be $n$ in order to cancel the
local gravitational anomaly on the worldsheet, by
making the left-moving central charge and right-moving central 
charge equal.  The left-moving fermions $\l\uu a, \cht\uu A$
do not have dynamical superpartners, so their contribution
to the left-moving central charge is $\hh$ per extra
dimension.

For $n$ extra dimensions, the extra central charge
for the right-movers is
\ee{
\Delta c = \Delta c \ll X + \Delta c \ll \psi = n + {n\over 2}
= {{3n}\over 2}.
}
The extra central charge for the left-movers is
\ee{
\Delta \tilde{c} 
= \Delta \tilde{c} \ll X + \tilde{c}
 \ll \cht = n + {n\over 2}
= {{3n}\over 2}.
}

The symmetry group of $\ho {+(n)}$ is $SO(32) \times
\lsq O(n) \times O(9+n,1)\rsq\ll +$.  The second factor
denotes
the subgroup $(g, g\pr)\in
O(n) \times O(9+n,1)$
with ${\rm det} [g] \cdot {\rm det} [g\pr] = 1$.

There are two discrete gauge symmetries on the
type $\ho +$ worldsheet.  The first
symmetry, $g\ll 1$, reverses the sign
of all 32 of the $\l\uu a$, just like the
$(-1)\uu{F_L}$ symmetry of type HO.  The second, $g\ll 2$,
reverses the sign of the right-moving
fermions $\psi\uu\m\ll +$, as well as the extra
left-moving fermions $\cht\uu A$.  $g\ll 2$ plays the role
of $(-1)\uu{F_{R_W}}$ in the supersymmetric HO theory.  In
particular it is a discrete R-symmetry; the supercharge
$Q\ll +$ is
odd under it.

The second eleven-dimensional heterotic string theory
we will discuss has the same field
content as $\ho +$, and differs
in its discrete worldsheet
gauge group.  In the second theory, we make only one
projection on fermions, retaining sectors of
even overall fermion number.

We call this theory type $\ho {+/}$.  The diagonal slash
refers to the fact that this theory is the same as
$\ho +$, only we take the projection on fermions which
corresponds to the diagonal modular invariant.

The theory generalizes in the obvious way to
the case of $n$ extra dimensions and $n$ extra current algebra
fermions.  In $10+n$ dimensions, the symmetry group
of the $\ho {+/}$ theory is $
\lsq O(32 + n) \times O(9+n,1)\rsq\ll +$.

\subsec{Spectrum of $\ho {+(1)}$}

The discrete charges of worldsheet fields in this
model are as follows:

\bf Table 1: \rm Discrete worldsheet gauge charge assignments in
model $\ho {{+(1)}}$.
\bigskip
\begintable
object | $g\ll 1\simeq (-1)^{F_{L_W}}$ | $g\ll 2
\simeq (-1)^{F_{R_W}}$ 
\elttt {3 pt}
$Q_+$                           | + | -   
 \elt
$X\uu\m$                        | + | +  
 \elt
$\psi\ll +\uu\m $               | + | -   
\elt
$\l\ll -\uu a$                  | - | +    
\elt
$\cht\ll -$                     | + | -

\endtable
\bigskip
\noindent
The $\simeq$ above indicates that $g\ll {1,2}$ act on
$\cht\ll -$ with the opposite of the sign with which they
act on all the other left-moving fields.

In order to calculate the spectrum, first we write the
super-Virasoro generators for the matter theory in terms
of oscillators.

\ee{
L\uu{mat.}\ll n - &A\uu{mat.}\d\ll{n,0} =
\cr
\hh \sum\ll{m} :~\a\uu\m\ll{n-m} \a\ll{\m m} ~:
+ \ff \sum\ll{r} (2r-m) &:~\psi\uu\m\ll{n-r} \psi\ll{\m r} ~:
+ i \left ( {\apr \over 2} \right )\uu\hh (n+1)
V\ll\m \a\uu\m\ll n
\cr
&
\cr
G\uu{mat.}\ll r = \sum\ll n \a\uu\m\ll n \psi\ll{\m~r-n}
+ i &\lrd {\apr\over 2} \rrd \uu\hh (2r+1)V\ll\m \psi\uu\m\ll r
\cr
&
\cr
\tilde{L}\uu{mat.}\ll n  - &\tilde{A}\uu{mat.} \d\ll{n,0} =
\cr
\hh \sum\ll{m} :~\tilde{\a}\uu\m\ll{n-m} \tilde{\a}\ll{\m m} ~:
+ \ff \sum\ll{r} &(2r-m) :~\l 
\uu a\ll{n-r} \l\uu a\ll r  ~: +
\cr 
+
\ff \sum\ll{r} (2r-m) :~\tilde{\chi} 
\ll{n-r} \tilde{\chi}\ll r  ~:
&+ i \left ( {\apr \over 2} \right )\uu\hh (n+1)
V\ll\m \a\uu\m\ll n
}
The linear dilaton term changes the central charge by
$\Delta c = 6 \apr V\ll\m V\uu\m,$
so in order to make the total central charge in the matter sector
equal to $(26, 15)$, we must take
\ee{
V\ll\m V\uu\m = - {1\over {4\apr}} \lrd D - 10 \rrd
}

According the standard rules, explained in
(\PolchinskiRQ, \PolchinskiRR), we compute the normal-ordering
contributions $A\uu{mat.}$ and $\tilde{A}\uu{mat.}$
to $\tilde{L}\uu{mat.}\ll 0$ and $L\uu{mat.}\ll 0$ in each
sector.  We also list the normal-ordering contributions
$\tilde{A}\uu{gh.}, A\uu{gh.}$
to $\tilde{L}\uu{gh.}\ll 0, L\ll 0\uu{gh.}$ from
the ghosts.  As in the usual heterotic
string, $\tilde{A}\uu{gh.}$  is always $-1$ and
$A\uu{gh.}$
depends only on the periodicity of the supercurrent
in each sector.

\bf Table 2: \rm Normal ordering contributions to
$\tilde{L}\ll 0$ and $L\ll 0$  in various sectors,
for the $\ho {+(1)}$ theory.
\bigskip
\begintable
$\matrix { & {\rm sector} \rightarrow  \cr {\rm field} &
\cr \downarrow & \cr & }$
| 1 | $g\ll 1$ | $g\ll 2$ | $g\ll 1 g\ll 2$ 
\elttt {3 pt}
$\tilde{b},\tilde{c},b,c,\b,\g$ | $(-1,-\hh)$ | $(-1,-\hh)$ | 
$(-1,-{5\over 8})$ | $(-1, -{5\over 8})$ 
\elttt {3 pt}
$X\uu\m$ | $(0,0)$  | $(0,0)$  
| $(0,0)$  | $(0,0)$  
 \elt
$\psi^\mu_+$ | $(0,0)$ | $(0,0)$ |
$(0,+{{11}\over{16}})$ | $(0,+{{11}\over{16}})$
 \elt
$\l\ll -\uu a$ | $(0,0)$ | $(+2,0)$
| $(0,0)$ | $(+2,0)$
\elt
$\cht\ll -$ | $(0,0)$ | $(0,0)$
| $(+{1\over{16}},0)$ | $(+{1\over{16}},0)$
\elttt {3 pt} 
$(\tilde{A}\uu{total}, A\uu{total})$|
 $(-1,-\hh)$ | $(+1 , -\hh)$ | $(-{{15}\over{16}},+{ 1\over
{16}})$ | $(+{{17}\over{16}}, +{1\over{16}})$
\endtable
\bigskip
\noindent

First we examine the lowest level with momentum
$k$ in the untwisted sector
$\left \abs  ~~~ \right \rangle\ll 1$.  
We need to act on the vacuum $\left \abs  k,0\right \rangle\ll 1$
with some object which is odd under the symmetries $g\ll 2$
and $g\ll 1 g\ll 2$, since
the effect of the superghosts is to make the vacuum odd
under these elements.
Since $\tilde{\chi}$ is odd under both and antiperiodic
in the untwisted sector, we can make the state
\ee{
\left \abs  T(k) \right \rangle \equiv
\tilde{\chi}\ll{-\hh} \left \abs  k,0\right \rangle\ll 1
}
which satisfies the physical state conditions if and only if
\ee{
k\sqd + 2 i (V\cdot k) = {2\over{\apr}}
}

Changing the $k$'s into $-i\pp$'s, this gives the linearized
EOM for the tachyon
\ee{
\pp\sqd T - 2 (\pp\phn) (\pp T) + {2\over{\apr}} T = 0,
}
which comes from the Lagrangian
\ee{
{1\over{\k\ll {11}\sqd}} \exp{-2\Phi} \sqrt{\abs  G \abs  } \lsq
-\hh (\nabla\ll\m T)(\nabla\uu\m T) + {1\over{\apr}} T\sqd \rsq
}

By a calculation which precisely follows those
of (\deAlwisPR, \MyersFV, \ChamseddineFG), the rest of the
states of the untwisted sector also appear in the action with their
usual kinetic terms, in addition to the potential for
the dilaton:
\ee{
{\cal L}\ll{\rm untwisted}\uu{\ho +}
=  {1\over{2\k\ll {11}\sqd}} \exp{-2\Phi} \sqrt{ \abs  G \abs  }\cdot
\lsq  \matrix {
- {{ 1}\over{\apr}} + 4 (\nabla\ll\m \Phi)(\nabla\uu\m \Phi)
\cr
\cr
-\hh (\nabla\ll\m T)(\nabla\uu\m T) + {1\over{\apr}} T\sqd 
\cr
\cr
+ {\bf R} - {1\over{12}} {\tilde{H}}\ll{\m\n\l}
 {\tilde{H}}\uu{\m\n\l} 
\cr
\cr
- {{\k\ll{11}\sqd}\over{2 g\sqd\ll{11}}}
{\rm Tr}\ll{\rm v} (F\ll{\m\n} F\uu{\m\n})
+ O(\apr)
}
\rsq
}
The 
subscript 'v' denotes a trace taken in the vector representation
of $SO(32)$ and the
tilded field strength $\tilde{H}$ is given by the sum
\ee{
\tilde{H} \equiv d B - {{\k\ll{11}\sqd}\over{g\ll{11}\sqd}}\o\ll 3,
}
of the curl of the B-field and a multiple of the Chern-Simons
form 
\ee{
\o\ll 3 \equiv {\rm Tr}\ll {\rm v} \lrd A \ww dA + {2\over 3} 
A\ww A\ww A \rrd.
}
Considering a certain class of tree-level solutions will serve to
fix ${{\k\ll {11}\sqd}\over{g\ll{11}\sqd}}$ in terms
of $\apr$, but we defer that analysis until later
in the section.  For now we continue to explore the spectrum
of the $\ho{+(1)}$ theory.

The sectors twisted by $g\ll 1$ and $g\ll 1\cdot g\ll 2$ are
less interesting for us, containing only massive 
bosons and fermions, in the sense that we will always use the
words 'massive' and 'massless' in this paper: their equations
of motion are those of fields with an explicit mass term
in the Lagrangian; we will not include such fields in
any of the effective actions we consider.

Remaining is the sector $g\ll 2$, which contains
massless fermions.

Let
$\left \abs  k,\a,0\right \rangle\ll{g\ll 2}$ be the twisted vacua, which 
transform in a representation (with basis labeled by $\a$) of
the Clifford algebra
\ee{
\{\psi\ll 0 \uu\m, \psi\ll 0\uu\n\} = \eta\uu{\m\n}
\llsk\llsk \{\psi\ll 0\uu\m , \cht\ll 0 \} = 0 \llsk\llsk 
\{\cht\ll 0 , \cht\ll 0\} = 1
}
generated by the zero modes of $\cht$ and $\psi\uu\m$.

The vacua are not level-matched and cannot be physical
states.  Nonetheless let us analyze their transformation
properties under the symmetries of the theory, since the physical
states which we build by acting with positive-frequency
oscillators will inherit gauge and Lorentz
quantum numbers from the ground states.

The Clifford algebra is generated by twelve elements whose total
signature is $(11,1)$.  The states transform as
spinors of $SO(11,1)$, though the dynamics do not respect this
symmetry since one of the eleven positive-signature generators
is the zero mode of a left-moving, rather than right-moving
fermion.

The properties of spinor representations of this algebra
are analyzed in an appendix.  The result is that, once
the GSO projection and the reality condition
are taken into account, is that the ground states
transform as a single Majorana spinor of $SO(10,1)$.

The ground state has $L\ll 0 - \tilde{L}\ll 0 = 1$,
so in order to make a level-matched state we must act
on the ground state $\left \abs  \a,k,0\right \rangle\ll{g\ll 2}$ with
a set of left-moving creation operators of total energy +1.
So our candidate physical states are
\ee{
\left \abs  \psi\ll \a\uu{[ab]} (k) \right \rangle &\equiv
\l\ll {-\hh}\uu a \l\ll{-\hh}\uu b \left \abs  \a,k,0\right \rangle\ll{g\ll 2},
\cr
\left \abs  \Psi\uu\m\ll\a (k) \right \rangle &\equiv
\tilde{\a}\uu\m\ll{-1} \left \abs  \a,k,0\right \rangle\ll{g\ll 2},~~~~~~~~
~~~~~~~~~{\rm and}
\cr
\left \abs  \th\ll\ald (k) \right \rangle &\equiv \cht\ll{-1} \left \abs  \ald,k,0\right \rangle
\ll{g\ll 2}
}

Since $g\ll 2$ is a Ramond sector,
the on-shell condition for all three comes from the $G\ll 0$
constraint; physical states satisfy the modified massless
Dirac equation
\ee{
(k +  i V )\ll\m \G\uu\m\ll{\a\b}
\left \abs  \psi\ll \b\uu{[ab]} (k) \right \rangle
= (k +  i V )\ll\m \G\uu\m\ll{\a\b} 
\left \abs  \Psi\uu\n\ll\b (k) \right \rangle = 
(k +  i V )\ll\m \G \uu\m\ll{\ald\bed} 
\left \abs  \th\ll\ald (k) \right \rangle = 0.
}
By 'NS sector' ('Ramond
sector') we will mean
a sector
with boundary conditions which make the supercurrent
even (odd).

Notice that the dilaton gradient appears in the combination
$k+ i V$ in these dispersion relations, rather than the
combination $k + 2 i V$ with which it appears in the
dispersion relations for NS sector fields.

These equations of motion can be obtained from the lagrangian
\ee{
{\cal L}\ll{fermi}\uu{\ho +} = 
 {i\over{2\k\ll {11}\sqd}} \exp{-2\Phi} \sqrt{ \abs  G
\abs  }
\lsq
 \Tr\ll v \lrd \psb \G\uu\m \nabla\ll\m \psi \rrd  
+ \thb \G\uu\m \nabla\ll\m \th + \bar{\Psi}
{}\uu\n  \G\uu\m
\nabla\ll\m \Psi\ll\n
\rsq
}

Despite the appearance of $k+ i V$ in the dispersion
relations, 
there is still the same
$\exp{-2\Phi}$ multiplying the kinetic term for the fermions
as for the massless bosons.  The combination $k+iV$
arises because there
is only a single derivative in the fermi
kinetic action; in the Euler-Lagrange equations,
there are two terms in which the derivative acts on
the fermions and only one term in which it acts on the dilaton.

The vector-spinor $\Psi\ll\a\uu\m$ can be
decomposed into a spin $1/2$ field and a spin-$3/2$ field
transforming according to a fermionic gauge invariance.

\subsec{Spectrum of the $\ho{+(1)/}$ theory}

The worldsheet theory of
$\ho{+/}$ has only one discrete gauge symmetry --
a projection onto even overall fermion number.
The left-moving fermions $\l\uu a$ have an index
which runs from one to $n+32$ and the $X\uu\m, \psi\uu\m$
have an index which runs from $0$ to $n+9$.

\bf Table 3: \rm Discrete worldsheet gauge charge assignments in
the 11D nonsusy heterotic theory.
\bigskip
\begintable
object |  $g\ll 1\equiv (-1)^{F_W}$  
\elttt {3 pt}
$Q_+$|- \elt $X^\mu$ |+ \elt $\psi^\mu_+$ |- \elt
$\l\ll -\uu a$                  | -      
\endtable
\bigskip
\noindent

This choice is modular invariant in any dimension, as long
the number of
current algebra fermions minus the number of spacetime
dimensions is equal to 22.  The continuous
part of the Lorentz group is $SO(n+9,1)$,
spontaneously broken only by the dilaton gradient; 
the continuous internal
gauge group is $SO(n+32)$. 

Specializing to the case of $\ho {+(1)/}$,
we calculate the normal ordering contributions to
the theory in its two sectors:

\bf Table 4:\rm Normal ordering contributions to
$\tilde{L}\ll 0$ and $L\ll 0$  in various sectors,
for the $\ho {+(1)/}$ theory.
\bigskip
\begintable
$\matrix { & {\rm sector} \rightarrow  \cr {\rm field} &
\cr \downarrow & \cr & }$
| 1 | $g\ll 1 $
\elttt {3 pt}
$\tilde{b},\tilde{c},b,c,\b,\g$ | $(-1,-\hh)$ | 
$(-1,-{5\over 8})$  
\elttt {3 pt}
$X\uu\m$ | $(0,0)$  | $(0,0)$  
 \elt
$\psi^\mu_+$ | $(0,0)$ | 
$(0,+{{11}\over{16}})$
 \elt
$\l\ll -\uu a$ | $(0,0)$ | $(+{{33}\over{16}},0)$
\elttt {3 pt} 
$(\tilde{A}\uu{total}, A\uu{total})$|
 $(-1,-\hh)$ |  $(+{{17}\over{16}}, +{1\over{16}})$
\endtable
\bigskip
\noindent

The analysis of the spectrum goes as in the $\ho{+(1)}$
theory, with three differences:

$\bullet{}$ The massless gauge
bosons $A\uu{[ab]}\ll\m$
obey the same equation of motion and transversality
condition as in $\ho{+(1)}$, but there are more of them since the
continuous gauge group is $SO(33)$.

$\bullet{}$ The tachyons $T\uu a$, obtained by acting with
$\l\uu a\ll{-\hh}$ on the untwisted vacuum, are more numerous
and transform in the vector representation of $SO(33)$.

$\bullet{}$ There are no
massless fermions at all.  

The action generating the free equations of motion for
the massless fields looks
exactly like the action for the untwisted sector of $\ho +$,
except that the tachyon is a vector instead of a singlet:

\ee{
{\cal L}\uu{\ho{+/}}
=  {1\over{2\k\ll {11}\sqd}} \exp{-2\Phi} \sqrt{\abs  G \abs  }\cdot
\lsq  \matrix {
- {1\over{\apr}} + 4 (\nabla\ll\m \Phi)(\nabla\uu\m \Phi)
\cr
\cr
-\hh (\nabla\ll\m T\uu a )(\nabla\uu\m T\uu a)
 + {1\over{\apr}} T\uu a T\uu a 
\cr
\cr
+ {\bf R} - {1\over{12}} {\tilde{H}}\ll{\m\n\l}
 {\tilde{H}}\uu{\m\n\l} 
\cr
\cr
- {{\k\ll{11}\sqd}\over{2 g\sqd\ll{11}}}
{\rm Tr}\ll{\rm v} (F\ll{\m\n} F\uu{\m\n})
+ O(\apr)
}
\rsq
}

The absence of massless spacetime fermions makes the
$\ho {+/}$ theory seem out of place in or discussion.
If our interest is in theories which can reach
supersymmetric critical
vacua by tachyon condensation, what is $\ho {+/}$ doing
here?  If the theory could relax to a supersymmetric
vacuum, where could the gravitini, dilatini, and gluini possibly
come from?

The premise of that question is correct: the $\ho{+/}$,
in its original noncompact version or compactified on any
smooth space, can never reach the supersymmetric HO theory by
tachyon condensation.  But we will show that
the $\ho{+(n)/}$ theory on spaces with certain orbifold
singularities of codimension $n$ can reach the
supersymmetric HO theory by tachyon condensation.
These spaces also have chiral fermions living
on the singularity.

\subsec{Generalization to arbitrary $n$}

At higher $n$, the $\ho +$ theory has
$10 + n$ spacetime embedding coordinates $X\uu\m$,
thirty-two current algebra fermions $\l\uu a$
and another $n$ current algebra ferions $\cht\uu A$.
The continuous symmetry group is
is $SO(9+n,1)\times SO(n) \times SO(32)$, where the first factor
rotates the $X\uu\m, \psi\uu\m$ coordinates, the
second factor rotates the $\cht\uu A$'s, and the third rotates
the $\l\uu a$'s.  The $SO(9,1)$ intercommutes in the usual
way with translations to make the ten-dimensional Poincar\'e
group, which is broken down spontaneously to the
nine-dimensional Poincar\'e group by the background
dilaton gradient.  The discrete gauge charges of the
worldsheet theory are given by a table that looks
essentially the same as in the case $n=1$; the only
difference is that there are more $\cht\uu A, X\uu\m,$
and $\psi\uu\m$.

\bf Table 5: \rm
Discrete worldsheet gauge charge assignments in
model $\ho {{+(n)}}$. 
\bigskip
\begintable
object | $g\ll 1\simeq (-1)^{F_{L_W}}$ | $g\ll 2
\simeq (-1)^{F_{R_W}}$ 
\elttt {3 pt}
$Q_+$                           | + | -   
 \elt
$X\uu\m$                        | + | +  
 \elt
$\psi\ll +\uu\m $               | + | -   
\elt
$\l\ll -\uu a$                  | - | +    
\elt
$\cht\ll -\uu A$                | + | -

\endtable
\bigskip
\noindent

The table of normal-ordering contributions to the
Virasoro generators $\tilde{L}\ll 0$ and $L\ll 0$ also
looks similar to the table for $\ho{+(1)}$,
though here there is an explicit $n$-dependence.

\bf Table 6: \rm
Normal ordering contributions to
$\tilde{L}\ll 0$ and $L\ll 0$  in various sectors,
for the $\ho {+(n)}$ theory. \rm
\bigskip
\begintable
$\matrix { & {\rm sector} \rightarrow  \cr {\rm field} &
\cr \downarrow & \cr & }$
| 1 | $g\ll 1$ | $g\ll 2$ | $g\ll 1 g\ll 2$ 
\elttt {3 pt}
$\tilde{b},\tilde{c},b,c,\b,\g$ | $(-1,-\hh)$ | $(-1,-\hh)$ | 
$(-1,-{5\over 8})$ | $(-1, -{5\over 8})$ 
\elttt {3 pt}
$X\uu\m$ | $(0,0)$  | $(0,0)$  
| $(0,0)$  | $(0,0)$  
 \elt
$\psi^\mu_+$ | $(0,0)$ | $(0,0)$ |
$(0,{{10 + n}\over{16}})$ | $(0,{{10 + n}\over{16}})$
 \elt
$\l\ll -\uu a$ | $(0,0)$ | $(+2,0)$
| $(0,0)$ | $(+2,0)$
\elt
$\cht\ll -$ | $(0,0)$ | $(0,0)$
| $({n\over{16}},0)$ | $({n\over{16}},0)$
\elttt {3 pt} 
$(\tilde{A}\uu{total}, A\uu{total})$|
 $(-1,-\hh)$ | $(+1 , -\hh)$ | $({{n - 16}\over{16}},{ n\over
{16}})$ | $({{n + 16}\over{16}}, {n\over{16}})$
\endtable
\bigskip
\noindent

When we examine the physical state
conditions, we use the fact that $V\ll\m V\uu\m = 
- {{D - 10}\over {4\apr}}$ and
find that the massless NS spectrum consists of
a graviton $G\ll{\m\n} - \eta\ll{\m\n}$, a B-field $B\ll{\m\n}$, a
dilaton $\Phi$, and a tachyon $T\uu A$ in the vector
representation of $SO(n)$.  The gauge transformations of
$G\ll{\m\n}$ and $B\ll{\m\n}$ are the usual
(dilaton-independent) ones,
and the equations of motion are the same as in $\ho{+(1)}$ when
given in terms of the dilaton gradient.  For example the
$B$-field obeys the equation of motion
\ee{
(\pp\uu\s - 2 V\uu\s) H\ll{\m\n\s} &= 0.
}

As for the Ramond states, the analysis likewise runs in
parallel.  The new feature comes from the fact that
the Ramond ground states, and hence all the massless
states in the $g\ll 2$-twisted sector, transform under $SO(n)$
in spinorial representations, because the $\cht\uu A$
have zero modes in this sector.  We derive the content
of the Ramond ground states in the appendix; the details of
the derivation
depend on the value of $n$ mod eight.

\bf Table 7: \rm Normal ordering contributions to
$\tilde{L}\ll 0$ and $L\ll 0$  in NS and R sectors,
for the $\ho {+(n)/}$ theory.
\bigskip
\begintable
$\matrix { & {\rm sector} \rightarrow  \cr {\rm field} &
\cr \downarrow & \cr & }$
| 1 | $g\ll 1 = (-1)\uu{F\ll W}$
\elttt {3 pt}
$\tilde{b},\tilde{c},b,c,\b,\g$ | $(-1,-\hh)$ | 
$(-1,-{5\over 8})$  
\elttt {3 pt}
$X\uu\m$ | $(0,0)$  | $(0,0)$  
 \elt
$\psi^\mu_+$ | $(0,0)$ | 
$(0,{{10 + n}\over{16}})$
 \elt
$\l\ll -\uu a$ | $(0,0)$ | $({{32 + n}\over{16}},0)$
\elttt {3 pt} 
$(\tilde{A}\uu{total}, A\uu{total})$|
 $(-1,-\hh)$ |  $({{16 + n}\over{16}}, {n\over{16}})$
\endtable
\bigskip
\noindent

\subsec{Smooth compactifications of $\ho +$}

We can put these theories on smooth spaces with
a curved metric and nonabelian gauge field strengths.

For untwisted perturbations,
the theories $\hop{} $ and $\hop /$ obey the same
one-loop
$\beta$-function equations as one another.  Those in turn are
the same equations as in the critical
heterotic string, except for the term depending on
the central charge:
\ee{
\b\uu{ G\ll{\m\n}} &= \apr {\bf R}\ll{\m\n}
+ 2\apr \nabla\ll\m \nabla\ll\n \Phi - {{\apr}\over 4}
H\ll{\m\r\s} H\ll\n{}\uu{\r\s}  - {{\apr \k\ll{11}\sqd}\over
{g\ll{11}\sqd}} \tr\ll v\lrd 
F\ll{\m\s} F\ll\n{}\uu\s \rrd + O(\apr\sqd)
\cr
\b\uu {B\ll{\m\n}} &= - {{\apr}\over 2} \nabla\uu\s H\ll{\m\n\s}
+ \apr (\nabla\uu\s \Phi) H\ll{\m\n\s} + O(\apr\sqd)
\cr
\b\uu{\Phi} &= {{D - 10}\over 4} - {{\apr}\over 2} \nabla\sqd
\Phi + \apr (\nabla\Phi)\sqd - {{\apr}\over{24}} H\sqd
- {{\apr \k\ll{11}\sqd}\over {4 g\ll{11}\sqd}}
\tr\ll v \lrd F\sqd \rrd + O(\apr\sqd)
\cr
\b\uu{A\ll\m\uu{[ab]}} &= {{2\apr\k\ll{11}\sqd}\over {g\ll{11}\sqd}}
\nabla\uu\n F\uu{[ab]}\ll{\m\n} - 
{{4 \apr\k\ll{11}\sqd}\over {g\ll{11}\sqd}}
(\pp\uu\n \Phi) F\ll{\m\n}\uu{[ab]} + O(\apr\sqd)
}

We only show the one-loop beta function here, but we can make
a statement about exact solutions to the $\b$-function
equations.

Suppose $n= \sum\ll i\uu m n\ll i$ and the
internal target manifold $\cx\ll n$ is a product
\ee{
\cx\ll n \equiv \cx\uu{(1)}\ll{n\ll 1} \times
\cx\uu{(2)}\ll{n\ll 2} \times \cdots 
\times \cx\uu{(m)}\ll{n\ll m}
}
of $m$ factors.  Suppose further that the vector bundle $V\ll n$
respects the product structure as well.  That is, it factorizes
into a product of vector bundles $V\ll {n\ll i}\uu{(i)}$ of rank
$n\ll i$, each over the space $\cx\ll {n\ll i}\uu{(i)}$.

If we choose the factors $(\cx\uu{(i)}\ll{n\ll i}, 
V\uu{(i)}\ll{n\ll i})$ such that the worldsheet sigma
model on each has an exact CFT limit, then it is clear
the sigma model on $\cx$ with bundle $V$ must be conformal
as well, since the product of CFT's is itself a CFT.

This gives a simple construction of an infinite class
of tree-level solutions to supercritical heterotic string
theory.  Since the rank of the vector bundle $V$
and the dimension of the internal space $\cx$ are both
equal to the same number $n$, these can be interpreted as
solutions either to $\ho +$ or to $\hod {+}$.

The existence of
this class of exact solutions serves to fix the coefficients
of interaction terms for the untwisted massless levels
in the effective action (3.9), (3.16).  In particular,
it determines~\PolchinskiRR
\ee{
{{\k\ll{11}\sqd}\over{g\ll{11}\sqd}} =
{{\k\ll{10}\sqd}\over{g\ll{10}\sqd}}
= {{\apr\over 4}} 
}

\subsec{Euler number and its generalization}

Real vector bundles of rank $n$ over orientable n-dimensional
maifolds with structure group $SO(n)$ can be classified according
to a certain topological invariant called the \it Euler number. \rm

Consider a section $\ct$ of the bundle $V$, given in a local basis
by $T\uu A(X), ~A = 1,\cdots,n$.  Generically $\ct$ will vanish
only
at isolated points on $\cx$.  Each vanishing point $x\ll 0\uu i$ 
for $T\uu A$ can be assigned a sign $\s\equiv \pm 1$ as
follows:
\ee{
\s\ll{ \abs\ll{x\uu i\ll0}}
\equiv \sgn \lsq {{dT\uu 1 \ww dT\uu 2 \ww\cdots\ww dT\uu n }
\over{dx\uu 1 \ww dx\uu 2 \ww\cdots \ww dx\uu n}} \rsq
_{x\uu i = x\uu i \ll 0}
= \sgn\det\lsq \matrix { \dt 1 1 & \dt 1 2 & \cdots & \dt 1 n \cr 
                                 & & & \cr
               \dt 2 1 & \dt 2 2 & \cdots & \dt 2 n \cr
               \cdots  & \cdots & \matrix { \cdots \cr
                                             \cdots \cr
                                             \cdots \cr }&\cdots \cr
               \dt n 1 & \dt n 2 & \cdots & \dt n n  } \rsq
_{x\uu i = x\uu i \ll 0}
}

The reader can satisfy herself that deformations of the
section can only created isolated zeroes in pairs with
opposite sign $\s$.  (A good introduction
to this and other aspects of the Euler number
for physicists can be found
in \BlauPM.)
Since the manifold is orientable and
the structure group of the vector bundle is $SO(n)$, the
sign $\s$ of a vanishing point is well defined under a change of
local coordinates and local basis for $V$.  Summing
$\s$ over vanishing points $x\uu i\ll 0$ gives a
topological invariant of the pair $(\cx,V)$.  This quantity
is called the \it Euler number \rm of the vector bundle $V$
over $\cx$.

The definition of
this quantity can be extended to the case where the structure group of the product $T\ll \cx \ot V$ of the tangent bundle and the bundle
$V$ has overall special orthogonal structure group
\ee{
\lsq O(n)\times O(n)\rsq \ll +,
}
but not
$SO(n) \times SO(n).$
The symmetry groups of the $\ho +$ and $\ho {+/}$ theories
are 
\ee{
\lsq O(n+9,1) \times O(n) \rsq \ll + \times SO(32)
}
and
\ee{
\lsq O(n+9,1) \times O(n+32) \rsq\ll +;
}
the compactifictions we consider have transition functions
in the subgroup $\lsq O(n) \times O(n) \rsq \ll +$,
embedded in the obvious way.  This means we can consider spaces
where the coordinate transition functions between patches of $\cx$
are orientation reversing, so long as the change of local basis
on $V$ has a compensating sign of $-1$ in its determinant,
between the same two patches.

Thus, the sum
\ee{
\chi[V] \equiv \sum\ll{x\ll 0\uu i} \s
}
over vanishing points of the section is a topological invariant
of the pair $(\cx, V)$. 

If we ignore the coordinates $X\uu{0-9}$ and the current
algebra fermions $\l\uu{1-32}$, as well as the ghosts
and antighosts, we can consider the $(0,1)$ supersymmetric sigma
model on $X$ with vector bundle $V$ in its own right.  This theory
by itself is still modular invariant and has only one
$Z\ll 2$ gauge symmetry, overall fermion number mod $2$.

In this theory, the Euler number is equal to the Witten index
of the quantum theory, in the case where $\cx$ is
compact and $(\cx, V)$ is smooth.

To see this, consider the theory at large volume, where it can
be studied semiclassically.  Then perturb the Lagrangian with a
relevant operator of the form
\ee{
\cl = Q\ll + \cdot (T\uu A(X) \cht\uu A\ll -),
}
which equals
\ee{
\sum\ll A (T\uu A(X))\sqd - i T\uu A{}\ll{,i} (X) \psi\uu i
\ll + \cht\uu A\ll - + A\uu {[AB]}\ll i T\uu B (X)
\psi\ll + \uu i \cht\ll - \uu A
}

The index $A$ ranges from 1 to $n \equiv {\rm dim}(\cx)$, so
the zeroes of the positive definite
worldsheet potential $T\uu A T\uu A$ will
be isolated for generic perturbations of this form.  At an
isolated zero, the mass matrix $T\uu A {}\ll{,i}$
is nondegenerate.  

What is the fermion number (or the fermion number mod 2)
of such a vacuum?  Changing the sign of the frequency of a 
fermionic oscillator changes the ground state of that
oscillator from $\left \abs   \uparrow  ~\right \rangle$ to
$\left \abs   \downarrow ~ \right \rangle$, and so changes the
fermion number of the
vacuum by one.  This means that $(-1)\uu F$ for a given vacuum,
semiclassically localized at a zero of the section $T\uu A$, 
is given by $\sgn \det \{T\uu A{}\ll{,i}\}$ evaluated at the
point where the section vanishes.  It follows
that the Witten index is given
by summing the quantity $\sgn \det  \{T\uu A{}\ll{,i}\}$
over all zeroes of the tachyon, so the index is equal to the
Euler number of $(\cx, V)$.
We call this the
\it generalized Euler number \rm of the vector bundle
$V$ over $\cx$.

In the next section we will see that sections of $V$ 
can be thought of as configurations of the tachyon,
and the generalized Euler number will be the number of
disconnected ten-dimensional universes into which the 
$10+n$-dimensional universe fragments in the
process of tachyon condensation.

\newsec{Tachyon condensation in $\ho +$ theories}

Both $\ho +$ and $\ho {+/}$ have bulk tachyons whose
behavior is much like the behavior of the
closed string tachyon in the bosonic and type zero
theories: a background for the tachyon gives rise to an
effective potential on the string worldsheet
and destroys the spacetime being probed by
the fundamental string.

However the $\ho +$ and $\ho{+/}$ theories differ from
the bosonic and type $0$ theories in that the latter have
gauge-neutral tachyons, whereas in the former all tachyons transform
nontrivially under continuous gauge
symmetries.  We will exploit this property to compactify
the theories to ten dimensions
in such a way that the ten dimensional effective theories can
have stable minima.  Indeed we will argue that the minima
are exactly the ten-dimensional supersymmetric vacua of
the standard HO string theory.

Attempts
to understand closed string tachyon condensation in the type 0 and
bosonic theories
include \HorowitzGN, \SuyamaAS, \StromingerFN.  In none
of these cases is the endpoint of tachyon dynamics a state
which is Poincar\'e-invariant and stable, much less
supersymmetric.

We can also perturb the theory by giving an expectation value
to the tachyon field.  The vertex operator of
the tachyon involves a single
current algebra fermion mode of frequency $\hh$ acting on
the untwisted
vacuum, either $\cht\uu A\ll{-\hh}$
in the $\ho +$ theory or $\l\uu a\ll{-\hh}$ in the $\ho {+/}$ theory.

\subsec{Dynamical \rm vs. \it kinematical tachyon condensation}

We want to show that our theories are
connected via tachyon condensation to 
the supersymmetric vacuum of the ten-dimensional
supersymmetric HO theory.

When we say 'connected' by tachyon condensation, we could mean
one of two things.  We could mean that there exists a
dynamical solution
\ee{
G\ll{\m\n}(X), T\uu A (X), \Phi(X)
}
to the equations of motion
of string theory which approaches some solution of the
supercritical theory -- for instance, the timelike linear dilaton
background -- as $X\ll 0 \to - \infty$, and to the 
ten-dimensional HO theory as $X\uu 0 \to + \infty$.  Such
an interpretation is appealing and possibly
right, but many problems would be involved in making it work,
or even in stating precisely what spacetime fields
such a solution would be described in terms of.

We will not consider the question of whether such solutions
exist.  In this paper, when we refer to connecting the supercritical
theory to the critical theory by tachyon condensation, we will
always mean only that the two theories can be connected by a
renormalization group flow on the string worldsheet, where
the relevant operator perturbing the supercritical worldsheet
is the one which appears in the vertex operator for the
corresponding tachyon.  We will think of this as a kind of
off-shell, or kinematical, tachyon condensation.

\subsec{Local models for tachyon condensation}

Let us now compare and contrast local models of tachyon
condensation in various unstable string theories.

\bf Open string tachyons in bosonic string theory \rm

In open bosonic string theory the relevant operator corresponding
to open string tachyon condensation is a boundary
cosmological constant:
\ee{
\Delta {\cal L}\ll{\rm worldsheet} =
E\ll{\rm boundary} \cdot \d(\partial \Sigma)
}
We can give this a dependence on the target space dimensions
as long as we keep the scale of the variation large enough
so that the contributions to the anomalous dimension
is small and the operator is still relevant:
\ee{
\Delta{\cal L}\ll{\rm worldsheet} = T(X)
\cdot \d(\partial \Sigma),
}
where $T(X)$ can be thought of as representing the
bosonic open string tachyon.  When $T(X) = 0$ the
worldsheet theory describes open strings propagating
freely in all of spacetime; there is a single
space-filling D-brane.  Turning on the tachyon
$T(X) = {\rm const.} > 0$ creates a potential energy
which suppresses boundaries in the path integral
and leaves behind a theory of closed strings only. Turning
on $T(X) = {\rm const} < 0$ creates a potential which
\it favors \rm boundaries; for this choice the string
is unstable to nucleation of endpoints and the path
integral is not bounded.

One can also turn on the tachyon inhomogeneously in
space.  The profile $T(X) = m\sqd X\sqd$ creates a potential
which in the infrared confines the string endpoints to
$X = 0$.  The tachyon profile is varying in what can
be thought of locally as a kind of kink solution.  It can
be extended to a true 'kink' profile globally
by letting 
\ee{
T\ll{\rm kink} (X) = 1 - {1\over{\cosh{X\over L}}},  
}
with $L$ some distance scale much larger than $\apr$.
But
the kink is nontopological; the value of
the tachyon $T(X)$ is the
same for $X\to + \infty$ as for $X \to - \infty$.  As
a result the kink is unstable; the zero of the potential
can be removed without altering the asymptotics of
the solution:
\ee{
T\ll{\rm kink} (X) \to 1 - {{(1 - T\ll{D-1})}\over{\cosh
{X\over L}}}
}
If we give $T\ll{D-1}$ a small positive value, the
profile $T\ll{\rm kink}(X)$ no longer has a zero anywhere,
and string endpoints are banned from all of space.
The notation $T\ll{D-1}$ is meant to highlight the fact
that the parameter $T\ll{D-1}$ functions as a relevant
operator which destroys the codimension-one D-brane
defined by the zero of the undeformed kink solution.

More precisely, the perturbation $T\ll{\rm kink}(X)$
flows
to the theory of a single codimension-one D-brane.  
That infrared theory has a relevant operator corresponding
to the open string tachyon on the D-brane.  That
relevant operator can be lifted to the ultraviolet,
and when lifted it becomes the parameter
$T\ll{D-1}$.

\bf Open string tachyons in type IIB string
theory \rm

One way to represent the theory of a space-filling
brane-antibrane pair is to let a free complex fermion
$\g \neq \g\dag$ live on each boundary of the worldsheet
$\HellermanCT, \HellermanBU$.
Then the two states $\left \abs ~\uparrow ~\right \rangle$
and $\left \abs ~\downarrow ~\right \rangle$ for the
fermion $\g$ at each boundary describe two possible
states of each endpoint.  The state
$\left \abs ~\uparrow ~\right \rangle$ can be interpreted as
an endpoint living on the brane, and the
state $\left \abs ~\downarrow ~\right \rangle$ can be interpreted
as the endpoint living on the antibrane.  Then the ground
state 
fermion number mod two of the brane-antibrane strings
and the fermion number mod two of the brane-brane strings
in the matter sector are what they should be: odd and even,
respectively.  This is why the brane-antibranes strings
are tachyonic: their oscillator ground state is already odd,
so they do not need a positive-frequency
fermionic oscillator to act on them in order to satisfy
the GSO condition.

The vertex operator of the open-string tachyon which annihlates
the brane-antibrane pair is made from the fermion $\g$ with
appropriate spatial and temporal dependence:
\ee{
\Delta {\cal L} = \d(\partial \Sigma) ~Q \cdot W,
\llsk & \llsk W \equiv T(X) \g + h.c.,
}
where $Q$ is a Hermitean supercharge and $\g$ transforms as a
Fermi multiplet:
\ee{
Q\cdot \g = F\llsk & \llsk Q \cdot F = \dot{\g}.
}
After we integrate out the auxiliary field, the bosonic potential
is $T(X)\sqd$.  Note that this is positive definite and symmetric
under $T\to - T$.  Since the fermion $\g$ is complex, $T$ is
also a complex function.  We can consider various possible
local forms of $T$.  If $T \sim X\ll 1 + o (X\sqd)$ near the
origin, then the brane is codimension one.  Eightbranes
in type IIB string theory are uncharged and unstable against decay,
so there should be an open string tachyon in the CFT to
which the theory with
this superpotential flows.

The relevant operator corresponding to the
open string tachyon can be lifted to the UV theory; it
corresponds to the operator $i\g - i \g\dag$.  Perturbing
the boundary superpotential with this operator, with
coefficient $\e$, destroys the brane.  The total
superpotential becomes
\ee{
W = \g (X\ll 1 + i \e) + {\rm h.c.},
}
and the bosonic potential is
\ee{
\abs X\ll 1 + i \e \abs\sqd,
}
which vanishes nowhere.

We can also choose the local form of the tachyon profile to be
\ee{
T = X\ll 1 + i X\ll 2 
}
This represents a stable, codimension-two brane in the type IIB
theory.  The bosonic potential is
\ee{
\abs X\ll 1 + i X\ll 2\abs \sqd
}
Any perturbation of the tachyon which is bounded
at infinity leaves intact the zero of the boundary potential.

\bf Closed string tachyons in bosonic string theory. \rm

The vertex operator for the closed string tachyon is just
the identity, dressed with spatial dependence.  A spatially
homogeneous tachyon perturbation is a cosmological
constant on the worldsheet $\Delta{\cal L} = 
{\rm const.} = E$; a spatially varying
tachyon is a potential for the $X\uu i$ fields
\ee{
\Delta {\cal L} = T(X).
}
Just as we did for the bosonic open string tachyon, we can choose
$T(X)$ to depend on a single spatial coordinate
$X\ll 1$ in such a way that it has a single quadratic minimum
at zero: 
\ee{
T(X) = T(X\ll 1) = \hh m\sqd X\ll 1\sqd.
}
If we began, for example,
with a supercritical bosonic string theory in 26 + n dimensions,
the IR
limit of this theory describes a bosonic string
in 26 + (n-1) dimensions.\foot{In order for this to be true, the
linear dilaton coefficient must also be renormalized.  This can
only happen through the dressing of this solution with
dependence on the time coordinate $X\uu 0$
dependence.  The effect cannot show up in the kinematical approach
we are taking to tachyon condensation.}
Just as in the bosonic theory, we can perturb the
tachyon profile by lifting its minimum away from zero:
\ee{
\d T = {\rm const.} = \e;
}
in the IR
this perturbation becomes the bosonic closed string tachyon
in 26 + (n-1) dimensions.

\bf Closed string tachyons in type $\ho {+(1)}$ string theory \rm

This theory has a single real tachyon $T$ which couples
to the worldsheet as
\ee{
\Delta W = \cht ~T(X).
}
The bosonic potential on the worldsheet is 
\ee{
T(X)\sqd.
}
Locally near $X\uu{10} = 0$ we can choose $T(X)$ to be of the form 
\ee{
T(X) \sim {m\over {\sqrt{2}}} X\uu{10},
}
and the resulting bosonic worldsheet potential is 
\ee{
\hh m\sqd (X\uu{10})\sqd
}
This has a single zero, at $X\uu{10} = 0$.  So the
endpoint of tachyon condensation is a single
universe with ten dimensions.  It is easy to see that
the ten-dimensional universe has no tachyons which lift to
deformations of the eleven-dimensional solution;
the persistence
of a zero of the potential is stable under small perturbations
to the superpotential.

\bf Closed string tachyons in type $\ho{+(n)}$ \rm

The theory has $n$ real tachyons which couple to the 
worldsheet as $\Delta W = \cht\uu A T\uu A(X)$.  
We can restrict our attention to profiles which break
the $SO(n)$ gauge symmetry and $SO(n)$ rotational symmetry
of the internal dimensions down to a diagonal subgroup:
\ee{
T\uu A = \sum\ll i \delta\uu {A+9, i} X\uu i \cdot f(r),
~~~~~~~~~~~~~r \equiv \sqrt{X\uu B X\uu B}. 
}
There is a single zero of the tachyon at the origin
and this zero persists under arbitrary small perturbations,
even those which break the symmetry.
At the level of a local model, we can just take $f$ to be
a constant, ${m\over{\sqrt{2}}}$.  The resulting
worldsheet interaction gives a mass term for $X\uu {10}$ through
$X\uu {9+n}$ and their superpartners, and also for all the
$\cht\uu A$.

\subsec{Descent relations for chiral fermion spectra}

We do not know how to
describe the dynamical change in the number of dimensions 
in any kind of effective field theory.  In the massless
bosonic sector, describing the decrease in the number 
of spacetime degrees of freedom may require
mechanisms not yet known.

For the case of massless fermions, the situation is much
easier to understand.  Upon condensing a tachyon
or set of $k$ tachyons inhomogeneously, in order to go from
$D$ dimensional string theory to $D - k$ dimensional string
theory, the tachyon defect which represents the lower-dimensional
universe can be thought of as an ordinary field theory
soliton.  The chiral fermion spectrum of the $D-k$ dimensional
string theory comes out right if we just treat the
computation of the fermion spectrum as we would for the case
of fermions localized to an ordinary gauge theory defect.

The only inputs necessary for this calculation are
the form of the Yukawa coupling between the higher-dimensional
tachyon and fermions, and the topology of the tachyon
profile, which we discussed earlier in the section.
We focus on the $SO(32)$ adjoint fermions and the case
$D = 10 + n, ~k = n$.
\vfill\eject

\bf n = 1 \rm

The coupling of the adjoint fermions $\psi\ll\a\uu{[ab]}$
to the tachyon
in the $\ho{+(1)}$ theory
is of the form
\ee{
\int d\uu{11} x ~ y~T~ \psb\uu{[ab]}\psi\uu{[ab]},
}
where $y$ is a Yukawa coupling which can be determined from
a three-point amplitude on the sphere.  Without the
tachyon $T$, no mass term is possible for the
$\psi\uu{[ab]}\ll\a$ field because parity is an exact 
symmetry of $\ho{+(1)}$.  Under spacetime reflections
across the plane $x\uu i = 0$, the fermions transform as
\ee{
\psi\to \G\uu i \psi.
} 
The eleven-dimensional gamma matrices are real and symmetric
(except for $\G\uu 0$ which is real and antisymmetric), and
so a mass term would transform under parity as
\ee{
M\psb\uu{[ab]} \psi\uu{[ab]} \to - M\psb\uu{[ab]}\psi\uu{[ab]}.
}
The tachyon $T$ is a pseudoscalar, so the Yukawa coupling
involving $T$ is allowed.  

It is interesting to see what happens to the fermion
spectrum when the tachyon is condensed inhomogeneously
in a kink solution.
First consider the local approximation to
the kink, where $T(X) = m X\uu{10}$.  Linearizing about this
background, the equation of motion for $\psi\uu{[ab]}\ll\a$
becomes
\ee{
\G\uu\m\ll{\a\b}\pp\ll\m\psi\uu{[ab]}\ll\b + m y X\uu{10}
\psi\ll\a = 0 
}
Consider a state which is a plane wave in the
$X\uu{0-9}$ directions.  In order for it to be
massless in the $X\uu M$ directions, $\psi$ must be
a zero mode of the operator $\pp\ll{10} + m y X\uu{10} \G\uu{10}$.
A zero mode of that operator
looks like
\ee{
\psi\ll\a\uu{[ab]}
 = \exp{- \hh my \hat{\G}\uu{10}\cdot (X\uu{10})\sqd}\ll{\a\b}
\psi\uu{(0)[ab]}{}\ll\b,
}
which is normalizable only if
\ee{
\G\uu{[10]} \psi\uu{(0)[ab]} = + \psi\uu{(0)[ab]}
}
so the massless adjoint fermions of the ten-dimensional
critical string theory have definite chirality as
10D spinors.

This result is entirely independent of the global form
of the tachyon profile; if $T(X\uu{10})$ is any
profile interpolating between two minima
$-T\ll 0$ and $+ T\ll 0$ of a generic double-well tachyon
potential, a zero mode of the linearized action
will have the form
\ee{
\psi\ll\a\uu{[ab]}
 = \exp{- y \int \ll 0\uu {X\uu{10}} ds~ T(s) \hat{\G}
\uu{10} } \ll{\a\b}\psi\uu{(0)[ab]}\ll\b.
}
The integral in the exponent goes as $T\ll 0 \abs X\uu{10}\abs$
for large $\abs X\uu{10}\abs$.  So the massless
$SO(32)$ adjoint fermions in the
ten-dimensional theory are exactly those with the chirality
$\G\uu{10}\psi\uu{[ab]}
 = {{T\ll 0}\over{\abs T\ll 0 \abs}} \psi\uu{[ab]}$.

%

\bf General n $>$ 1 \rm

Instead of dealing individually with all eight values of
$n$ mod 8, we will use a trick to treat them all
uniformly.  We will simply treat the spinors $\psi$ as
complex Dirac spinors of SO(n) and SO(n+9,1) for the purposes
of solving for the zero modes;
the reality and GSO conditions can be imposed afterwards.
Since the linear operators multiplying the $\G\uu\m$
and $\g\uu A$ matrices are all real, the zero mode
equation commutes with the imposition of a reality condition.
And since the $\G\uu \m$ and $\g\uu A$ matrices
anticommute with the overall chirality operator $\G\g$,
the zero mode equation respects the
GSO condition as well.

So let $\psi$ be a complex Dirac spinor of $SO(9+n,1)$
and of $SO(n)$.  In terms of this object the coupling
to the tachyon in $10+n$ dimensions can be written as
\ee{
\int d\uu{10+n} x ~ i \psb \G\uu\m \nabla\ll\m \psi
+ y T\uu A \psb \g\uu A \psi
}

The linearized
equation of motion for ten-dimensional massless fermion modes is
\ee{
i ~\G\uu i \pp\ll i \psi + y \g\uu A T\uu A \psi = 0,
}
where the sum over $i$ is taken to run from $10$ to $9+n$.
For the local model, we can take $T\uu A = {m\over{\sqrt{2}}}
x\uu{A+9}$.  In that case, the operators $\CO\ll{A}
\equiv \G\uu {A+9} \pp\ll{A+9} - {{iym}\over{\sqrt{2}}}
x\uu {A+9}  \g\uu A$
anticommute with one another, and annihilate $\psi$.  For
$ym > 0$, this means all spinor components are zero except
those with eigenvalue $+1$ under $-i\G\uu{A+9} \g\uu A$.
This cuts down the number of spinor components by a factor of
$2\uu n$, exactly the right number to give a single massless
spinor in ten dimensions.  This applies to all the $\ho +$
massless fermion states.

\subsec{Baby universes from tachyon condensation}

In the subsection above, we have seen that
sections $T\uu A$ of the vector bundle $V$
(with sufficiently slow spatial variation)
correspond to relevant
perturbations of the worldsheet action. These in turn
can be thought of as off-shell configurations for
the tachyon.

In a compact space this tells us something
interesting about the endpoint of
tachyon condensation.  If $\chi$ is the
Euler number of the vector bundle $V$ over our
compact space $\cx$, then there are at least
$\chi$ zero energy worldsheet vacua.  Since this property
is a result of worldsheet supersymmetry it
is entirely independent of $\apr$ corrections to the 2D couplings.
Generically these vacua are isolated from one another, 
and so we are led to expect that dynamical
tachyon condensation
will lead to a set of at least $\chi[V]$ disconnected
universes.

For pairs $(V, \cx)$ with $\chi = 0$, the 10-dimensional tachyons
have the same property as the bosonic and type zero tachyons:
condensing them in a generic way does not lead to any
conformal two-dimensional worldsheet theory.  For $\chi = 0$
the effective tachyon potential in 10 dimensions does not have
a stable minimum describing a perturbative string theory.

For $(V, \cx)$ with $\chi \geq 2,$ condensing the tachyon 
leads to a worldsheet theory with multiple minima, and the
IR theory describes strings propagating in several disconnected
universes.  The consistency of an evolution from a single
connected universe to multiple disconnected universes in
a quantum theory is unclear.  At best, such an evolution
would raise
difficult questions about unitarity and the flow of information.
In addition, it would not seem possible to describe
tachyon condensation for $\chi \geq 2$ with any sort of
a conventional ten-dimensional effective field theory.
The fluctuations about the
stable vacuum would have to include more than one
ten-dimensional graviton state.

Only for the case $\chi = 1$ is it possible
to describe tachyon condensation to
a supersymmetric minimum in terms of ten-dimensional
effective field theory in a single
universe.  Therefore we should
restrict our attention to pairs $(V,\cx)$ of Euler number 
$\chi = 1$ if we want closed string tachyon
condensation to be described by conventional ten-dimensional
effective dynamics with a supersymmetric vacuum.

Pairs $(V,\cx)$ with $\chi = 1$ are rare.  One example
is the product
\ee{
\cx  = (\IR\IP\ll 2)\uu k
}
of $k$ real projective planes, with $V$ being the tangent
bundle to $\cx$.  We know of no
compact examples at all in which $\cx$ is
smooth and Ricci-flat.  Because of that, we will not pursue tachyon
condensation in $\ho +$ theories beyond the end of this section.

Some caution must be applied
when using the intuition with which we have
argued for the evolution to disconnected universes
when $\chi \geq 2$.
Our counting of baby universes
relies on an \it adiabatic
approximation\rm , in which time evolution is accurately
descibed by a 2D field theory whose couplings evolve with
$X\uu 0$.  Such an approximation may not be valid in
an $n+1$-dimensional background in which
tachyons are condensing dynamically.

For one thing, the characteristic wavelength of the tachyon
is $\sqrt{\apr}$, and so the metric of an $n+1$-dimensional
solution will likely have variation on string scale
due to the backreaction of $T\uu A$ on the metric.
As a result, $O(\apr)$ corrections to tbe $\b$-function
equations cannot be ignored.  Though we have argued that 
$O(\apr)$ corrections to the sigma model couplings cannot
change the minimum number of baby universes at $X\uu 0
\to \infty$, it is still possible that the $\apr$-corrections
will sum up in such a way that the 2D theory undergoes a
transition, past which it cannot be described by a
sigma model at all.

More surprisingly, the adiabatic approximation can break down
by another mechanism
even when all curvatures and gradients are arbitrarily
small.  This is best illustrated by an example.  Let $\cx$ be
a smooth orientable
space of dimension $n = 6$ and let $V$ be the tangent
bundle of $\cx$.  Then let $(\cx\pr, V\pr)$ be another
such pair with Euler number $\chi\pr$.

There exists
a smooth Euclidean space $Y\ll 7$ with boundary, such
that the oriented boundary $\pp Y\ll 7$ of $Y\ll 7$ is
equal to the disconnected sum of $\cx$ and $\cx\pr$.
This fact follows from the triviality of the \it oriented
cobordism group \rm in dimension six.  (For an introduction
to this set of concepts, see for example \Milnor.)

If we are allowed to
think of $Y\ll 7$ as an 'off-shell' configuration
contributing to the Euclidean path integral of string
theory, then there is no selection rule
or principle of continuity which forbids smooth
histories in which pairs $(\cx , V)$
change their Euler number.  Such histories can have
arbitrarily small curvature, so $\apr$-corrections to
the effective action do not serve to suppress them.

The oriented cobordism groups are trivial or cyclic
in many dimensions other than six \Milnor.
Taking into account the necessity
of defining spinors consistently
on the interpolating manifold $Y$ does provide some
additional selection rules \Anderson, but even these
fail to suppress most processes which would violate the
Euler number.

\newsec{A duality between critical and supercritical
string theories}

\subsec{$\ho {+(1)/}$ on spaces with boundary}

Smooth compactifications in the $\ho {+/}$ theory will not
be of interest to us.

Instead we would like to study the $\ho {+/}$ theory on
orbifolds.  Specifically, we want to study the behavior of
$\ho{+(n)/}$ near fixed loci of $Z\ll 2$
orbifold actions which reflect
$n$ of the spatial coordinates.  We will call this orbifold
group element $g\ll 2$.

Start with the case $n=1$.  We will orbifold $\IR\uu {10,1}$ by
reflection of the tenth spatial coordinate $X\uu{10}\to
-X\uu{10}$.  To preserve $(0,1)$ SUSY we must also act on
$\psi\uu{10}$ as $\psi\uu{10}\to - \psi\uu{10}$.

This
action in itself would lead to an orbifold without a modular
invariant partition function.  The requirement of modular
invariance is satisfied for an orbifold which is level-matched
in the twisted sector.  Since every twisted fermion contributes
$+{1\over{16}}$ to the ground state energy, it is sufficient to
require that the number of left-moving fermions odd under
each $Z_2$ differ from
the number of
right-moving fermions odd under the same $Z_2$ by a
multiple of $16$.  

For the case $n=1$
there is exactly one right-moving fermion $\psi\uu{10}$ which is
odd under the orbifold action $g\ll 2$, so we can get a
modular-invariant theory by acting with a minus sign on
all 33 of the left-moving fermions $\l\uu a$ with $g\ll 2$.  Below
we list the transformations of all the objects in the theory
under the two new elements of the discrete gauge group.

\bf Table 8: \rm Charges of worldsheet fields under $g\ll 1$
and $g\ll 1 g\ll 2$ in $\ho{+(1)}$ on
the half-line $\IR / Z\ll 2$.  $g\ll 1$ is the worldsheet
fermion number mod two symmetry $(-1)\uu{F_W}$ and
$g\ll 2$ is a reflection of the eleventh dimension, along with
an inversion of all 33 $\l\uu a$ fermions.  
\bigskip
\begintable
$\matrix { & \cr & {\rm orbifold~group} \cr
& {\rm element} \rightarrow  \cr {\rm object} &
\cr \downarrow & \cr & }$
| $g\ll 2$ | $g\ll 1 g\ll 2$
\elttt {3 pt}
$Q\ll + ~~~~~~~~~~~~~~~~~~~~$ | + | -  
\elt
$X\uu{0-9}~~~~~~~~~~~~~~~~~~~~$ | + | +  
\elt 
$X\uu{10}~~~~~~~~~~~~~~~~~~~~$ | - | -
\elt
$\psi^{0-9}_+~~~~~~~~~~~~~~~~~~~~$ | + | -
 \elt
$\psi\uu{10}\ll + ~~~~~~~~~~~~~~~~~~~~$ |- | +
\elt
$\l\ll -\uu a ~~~~~~~~~~~~~~~~~~~~$ | - | +
\endtable
\bigskip
\noindent

Every element of the orbifold group is of order 2.
In the sector twisted by an element $h$, each free
left-moving field
odd under $h$, bose or fermi, contributes $+{1\over{16}}$
to the ground state value of $\tilde{L}\ll 0$ in that sector;
the same applies to right-moving fields and their contributions
to the ground state value of $L\ll 0$.  The ghosts and
superghosts also make their usual contributions to the
ground states of NS and R sectors.

\vskip .5in
\bf Table 9: \rm Ground state contributions to $\tilde{L}\ll 0$
and $L\ll 0$ in twisted sectors of $\ho {+(1)/}$ on
the half-line $\IR / Z\ll 2$.
\bigskip
\begintable
$\matrix { & \cr
& {\rm sector} \rightarrow  \cr {\rm object} &
\cr \downarrow & \cr & }$
| $g\ll 2$ | $g\ll 1 g\ll 2$
\elttt {3 pt}
$\tilde{b},\tilde{c},b,c,\b,\g$ | $(-1,-\hh)$ | 
$(-1,-{5\over 8})$  
\elttt {3 pt}
$X\uu{0-9}$ | $(0,0)$  | $(0,0)$  
 \elt
$X\uu{10}$  | $({1\over{16}}, {1\over{16}} )$ |
$({1\over{16}}, {1\over{16}} )$
\elt
$\psi^{0-9}_+$ | $(0,0)$ | 
$(0,{5\over 8})$
\elt
$\psi^{10}_+$ | $(0,{1\over{16}})$ | 
$(0,0)$
 \elt
$\l\ll -\uu a$ | $({{33}\over{16}},0)$ | $(0,0)$
\elttt {3 pt} 
$(\tilde{A}\uu{total}, A\uu{total})$|
 $(+{9\over 8},-{3\over 8})$ |  
$(-{{15}\over{16}}, +{1\over{16}})$
\endtable
\bigskip
\noindent

The sector $g\ll 2$ will not be of interest to us; it contains only
massive fields, due to the large ground state energy contributed
by the periodic $\l\uu a$.  We will focus on the sector
$g\ll 1 g\ll 2$, which contains states which are massless
in the sense of (\supercrit).

The coordinate $X\uu {10}$ is odd under $g\ll 1 g\ll 2$, so the
states in this sector are localized to the fixed plane $X\uu{10}
= 0$.  The fermions $\psi\uu{0-9}$ are also odd under
$g\ll 1 g\ll 2$, so the states are spinors of $SO(9,1)$.
The fermion $\psi\uu{10}$ is untwisted, so these spinor states are 
\it not \rm representations of the full $SO(10,1)$, only of
the $SO(9,1)$ which acts on the fixed plane.  The supercharge
$Q\ll +$ is odd under $g\ll 1 g\ll 2$, so to preserve modular
invariance we insert a $(-1)$ into the partition function in
this sector; as a result these states are spacetime fermions.

Label the oscillator vacua in this sector by
\ee{
\left \abs  (k\ll 0,\cdots,k\ll 9),\a,0\right \rangle\ll{g\ll 1 g\ll 2}
} 
and
\ee{
\left \abs  (k\ll 0,\cdots,k\ll 9),\ald,0\right \rangle\ll{g\ll 1 g\ll 2}
} 
where $\a$ and $\ald$ are parametrize bases for positive-
and negative-chirality spinors of $SO(9,1)$, respectively.
As explained in the appendix, fermion zero modes
$\psi\uu M\ll 0 ~(M = 0,\cdots,9)$ act on the oscillator
vacua as
\ee{
\psi\uu M\ll 0 \left \abs  k\uu N,\a,0\right \rangle\ll{g\ll 1 g\ll 2}
&= {1\over{\sqrt{2}}} \sum\ll\ald \G\uu M \ll{\a\ald} 
 \left \abs  k\uu N,\ald
,0\right \rangle\ll{g\ll 1 g\ll 2}
\cr
&{\rm and}
\cr
\psi\uu M\ll 0 \left \abs  k\uu N,\ald,0\right \rangle\ll{g\ll 1 g\ll 2}
&= {1\over{\sqrt{2}}} \sum\ll\a \G\uu M \ll{\ald\a} 
 \left \abs  k\uu N,\a
,0\right \rangle\ll{g\ll 1 g\ll 2}
}
where the $\G\uu M\ll{\a\ald}$
and $\G\uu M\ll{\ald\a}$
are real $16\times 16$ matrices satisfying
the Dirac algebra
\ee{
\G\uu M\ll{\a\ald} \G\uu N\ll{\ald\b} + (M\leftrightarrow N)
= 2 \d\ll{\a\b}
\cr
\G\uu M\ll{\ald \a} \G\uu N\ll{\a\bed} + 
(M\leftrightarrow N) = 2 \d\ll{\ald\bed}
}

In order to obtain a level-matched state we need to act on the
oscillator vacua with a set of raising operators for the
left-moving oscillators, of total
energy $+1$.  There are several ways to do this.

Acting with $\at\ll{-1}\uu M$ gives us a 10-dimensional
Majorana-Weyl
vector-spinor field of each chirality:
\ee{
\Psi\ll\a\uu M &\leftrightarrow \at\ll{-1}\uu M \left \abs  k,\a, 0\right \rangle
\ll{g\ll 1 g\ll 2},
\cr
&{\rm and}
\cr
\Psi\ll\ald\uu M &\leftrightarrow \at\ll{-1}\uu M \left \abs  k,\ald, 0\right \rangle
\ll{g\ll 1 g\ll 2}.
}
Acting with two $\l\uu a\ll{-\hh}$ operators gives us
a ten-dimensional Majorana-Weyl spinor in the adjoint of
SO(32):
\ee{
\psi\uu{[ab]}\ll\a \leftrightarrow
\l\ll{-\hh}\uu a \l\ll{-\hh}\uu b \left \abs   k,\a,0\right \rangle\ll
{g\ll 1 g\ll 2} ,
}
and similarly for $\ald$.
The oscillators $\at\uu{10}$ are also half-integrally
moded in this sector like the $\l\uu a$, since $X\uu{10}$ is
odd under $g\ll 1 g\ll 2$.  So we can make another set
of states by acting on the oscillator vacua with
one $\l\uu a$ raising operator and one $\at\uu{10}$ raising 
operator:
\ee{
\ups\uu{a}\ll\a \leftrightarrow \l\uu a \ll{-\hh} \at\uu{10}
\ll{-\hh} \left \abs   k,\a,0\right \rangle\ll
{g\ll 1 g\ll 2} ,
}
and similarly for $\ald$.  We can also make a level-matched
state by acting with two $\a\uu{10}$ raising operators:
\ee{
\theta\ll\a \leftrightarrow  
\at\uu{10}\ll{-\hh} 
\at\uu{10}\ll{-\hh} \left \abs   k,\a,0\right \rangle\ll {g\ll 1 g\ll 2} ,
}
and similarly for $\ald$.

Having added twisted sectors, we now restrict to
gauge-invariant states.
In the untwisted sector, physical states
must be even under the 
combination of $X\uu{10} \to - X\uu{10}$ and
$\l\uu a \to - \l\uu a$.  For scalars
this means that states transforming
as odd-rank tensors under $SO(33)$ must vanish
at the fixed plane $X\uu{10} = 0$,
and states transforming as even-rank tensors must
have vanishing normal derivative at the fixed plane.
In particular, this means that the tachyon $T\uu a$
has the Dirichlet boundary condition $T\uu a = 0$
at $x\uu{10} = 0$.  We shall see that this is quite important,
as it will have the effect of stabilizing the system of localized
modes at the boundary when the tachyon is condensed; the boundary
condition for the tachyon prevents the bulk instability from
reaching the boundary.  In this sense, the system 
described in this section is the precise opposite of that described
in \AdamsSV, which describes a stable string theory with a
tachyon localized at an orbifold fixed plane.

Tensors get an extra minus sign under $g\ll 2$ and $g\ll 1 g\ll 2$
for each index along the $X\uu {10}$ direction.  We list the
resulting boundary conditions for bulk fields in a table below:
\bigskip
\bf Table 10: \rm Boundary conditions for bulk fields at
an $\IR\uu n / Z\ll 2$ orbifold singularity of stable
type in type $\ho{+(1)/}$ string theory.
\bigskip 
\begintable
Bulk field | Boundary condition at $x\uu {10} = 0$
\elttt {3 pt}
$T\uu a$ | Dirichlet 
\elt
$G\ll{MN} , B\ll{MN} $ | Neumann 
\elt
$G\ll{M~10}, B\ll{M~10}$ | Dirichlet 
\elt
$G\ll{10~10}$ | Neumann
\elt
$\Phi$ | Neumann
\elt
$A\ll M\uu{[ab]}$ | Neumann 
\elt
$A\ll{10}\uu{[ab]}$ | Dirichlet 
\endtable

In the $g\ll 1 g\ll 2$-twisted sector, the GSO projection has
the effect of selecting one of the two chiralities of
Majorana-Weyl spinor, for each state of the positive-frequency
oscillators.  Since $\at\uu M\ll{-1}, ~\l\uu a \ll{-\hh} 
\l\uu b\ll{-\hh},$ and $\at\uu{10}\ll{-\hh} \at\uu{10}\ll{-\hh}$
are invariant under the orbifold group, the corresponding spinors
all have the same 10D chirality, say left-handed.  $\at\uu{10}
\ll{-\hh} \l\uu a\ll{-\hh}$ is even under $g\ll 2$ but odd
under $g\ll 1$, so it has the same quantum numbers as $\psi\uu M$.
This means the state obtained by acting with $\at\uu{10}
\ll{-\hh} \l\uu a\ll{-\hh}$ can be made gauge invariant by acting
with a fermion zero mode $\psi\uu M\ll 0$ or equivalently 
choosing the opposite chirality for these spinor states.  So
for massless boundary fermion states, the effect of the GSO
projection is to eliminate $\Psi\uu M\ll\ald, \psi\uu{[ab]}
\ll\ald, \th\ll\ald$ and $\ups\uu a\ll\a,$ and to retain
$\Psi\uu M\ll\a, \psi\uu{[ab]}
\ll\a, \th\ll\a$ and $\ups\uu a\ll\ald.$

The astute reader may recognize this pattern of
representations for the gauge group SO(33); it is the same
pattern which occurs in the chiral fermion spectrum of type
I string theory in ten dimensions, to which a single additional
D9-brane and anti-D9-brane have been added~\SchwarzSF.

The extra D9-brane enhances the gauge group to SO(33), and the
single $\overline{{\rm D9}}$-brane does not give rise to any
gauge symmetry.  In this theory there is a left-handed
vector-spinor $\Psh\uu M\ll\a$ coming from the
closed string sector; a left-handed
adjoint $\psh\uu{[ab]}\ll\a$ coming
from the D9-D9 open strings; a right-handed
$SO(33)$ vector $\upsh\uu a\ll\ald$, coming from
the D9-$\overline{{\rm D9}}$ open strings; and a left-handed
singlet $\thh\ll\a$ coming from the
$\overline{{\rm D9}}$-$\overline{{\rm D9}}$ open strings.

\subsec{Generalization to higher $n$}

The agreement between the gauge group and chiral fermion
spectrum of the type I+D9+$\overline{\rm D9}$
and $\ho{+(n)/}$ theories persists for higher $n$, where
the number of brane-antibrane pairs added to the type I theory
is $n$, and the $\ho{+(n)/}$ theory has a codimension-$n$
fixed locus of a $Z_2$ involution.  In this subsection we describe
the $\ho{+(n)/}$
orbifold for and its spectrum for general $n$.  Our description
is abbreviated, as it runs in parallel to the $n=1$ case.

Starting with the $10+n$-dimensional $\ho{+(n)/}$ theory, 
which has continuous gauge group $SO(32+n)$, we separate the
coordinates into two groups, which we label $X\uu M, ~M = 0,\cdots,
9$ and $Y\uu s, ~s = 1,\cdots ,n$.  Our $Z\ll 2$ orbifold action
inverts all the coordinates $Y\uu s$ and simultaneously acts
with a minus sign on all $32+n$ $\l\uu a$-fields.

The calculation of the massless fermion spectrum 
in the $\ho{+(n)/}$ theory goes through as for $n=1$:

\vskip 3in
\bf Table 11: \rm Chiral fermion spectrum
of $\ho{+(n)/}$ at an $\IR\uu n / Z\ll 2$
orbifold singularity of stable type.
\bigskip
\begintable
state | spacetime field | $SO(32+n)$ rep. |
$SO(n)$ rep. | $\matrix { ~\cr SO(9,1) \cr {\rm spinor} \cr
{\rm chirality} \cr ~} $
\elttt {3 pt}
$\at \uu M \ll{-1}  \left \abs  k\uu N, \a \right \rangle\ll{g\ll 1 g\ll 2}$ |
$ \Psi\uu M\ll\a$| \bf 1\rm | \bf 1\rm |
$\matrix { ~\cr
+
~\cr
{\rm and } \cr + \ll {({\rm spin}~{3\over 2})} \cr~
}$
\elt
$\l\uu a \ll{-\hh} \l\uu b\ll{-\hh}
  \left \abs  k\uu N, \a \right \rangle\ll{g\ll 1 g\ll 2}$ |
$ \psi\uu {[ab]}\ll\a$| $\Lambda\sqd$[\bf 32+n]\rm | \bf 1\rm |
$\matrix{~\cr +\cr ~}$
\elt
$\at\uu{s}\ll{-\hh} \l\uu a\ll{-\hh}
  \left \abs  k\uu N, \ald \right \rangle\ll{g\ll 1 g\ll 2}$ |
$ \ups\uu {a\abs s}\ll\ald$| \bf 32+n\rm | \bf n \rm | 
$\matrix{~\cr -\cr ~}$
\elt
$\at\uu{s}\ll{-\hh} \at\uu t\ll{-\hh}
  \left \abs  k\uu N, \a \right \rangle\ll{g\ll 1 g\ll 2}$ |
$ \th\uu {(st)}\ll\a$| \bf 1\rm | Sym$\sqd$[\bf n\rm] | 
$\matrix{~\cr +\cr ~}$
\endtable

For arbitrary $n$,
the spectrum of massless fermions is the same
as that of
the type I theory with $n$ extra D9-$\overline{\rm D9}$ pairs
added~\SchwarzSF.

In the type I theory there are also tachyons
$\hat{T}\uu{a\abs s}$ which transform in
the bifundamental of $SO(32+n)\times SO(n)$.  They are
the lowest lying states in the NS sector of the
D9-$\overline{\rm D9}$ open strings.
In the twisted sector of the $\ho{+(n)/}$ theory, the role
of the open string tachyon is played by the normal derivative
$\nabla\ll s T\uu a\ll{\abs\ll {Y\uu t = 0}} = \pp\ll s T\uu a
\ll{\abs\ll{Y\uu t = 0}}$ of the bulk tachyon
$T\uu a$ at the orbifold fixed locus $Y\uu t = 0$.
It is the normal derivative which participates in the most
relevant coupling between the boundary fermions and the bulk
tachyon:
\ee{
\cl \ll{{\rm INT}}\uu{(9+1)} = 
\upsilon\ll 1 (\pp\ll s T\uu b)\ll{\abs\ll {Y\uu u = 0}}
  \psb\uu{[ab]} \ups\uu{a\abs s} +
\upsilon\ll 2 (\pp\ll t T\uu a) 
\ll{\abs\ll {Y\uu u = 0}}
\thb\uu{(st)} \ups\uu{a\abs s} 
}
where $\upsilon\ll 1$ and $\upsilon\ll 2$ are Yukawa couplings
which can be determined by a three-point
computation in the tree-level $\ho{+(n)/}$ theory.

The same couplings occur in the
type I+$n$
D9+$n~\overline{\rm D9}$
system (with different coefficients), with the correspondence
\ee{
\psi\uu{[ab]} &\to \psh\uu{[ab]}
\cr
\ups\uu{s,a} &\to \upsh\uu{s,a} 
\cr
\th\uu{(st)} &\to \thh\uu{(st)}
\cr
{{\pp T\uu a}\over{\pp Y\uu s}}\ll{\abs\ll{Y\uu u = 0}}
 &\to \hat{T}\uu{a\abs s}
}
Hatted quantities refer to type I fields.

\ifig\sduality{
An S-duality between critical open string theories and
noncritical closed string theories.  The $\IR^n / Z_2$
singularity is of the stable, gauge-invariant
type type we have discussed, at which all tachyons have
Dirichlet boundary conditions.  The S-duality commutes
with tachyon condensation.
As we shall show in the next section, this diagram is
a simplification of the true phase structure.
}
{\epsfxsize3.0in\epsfbox{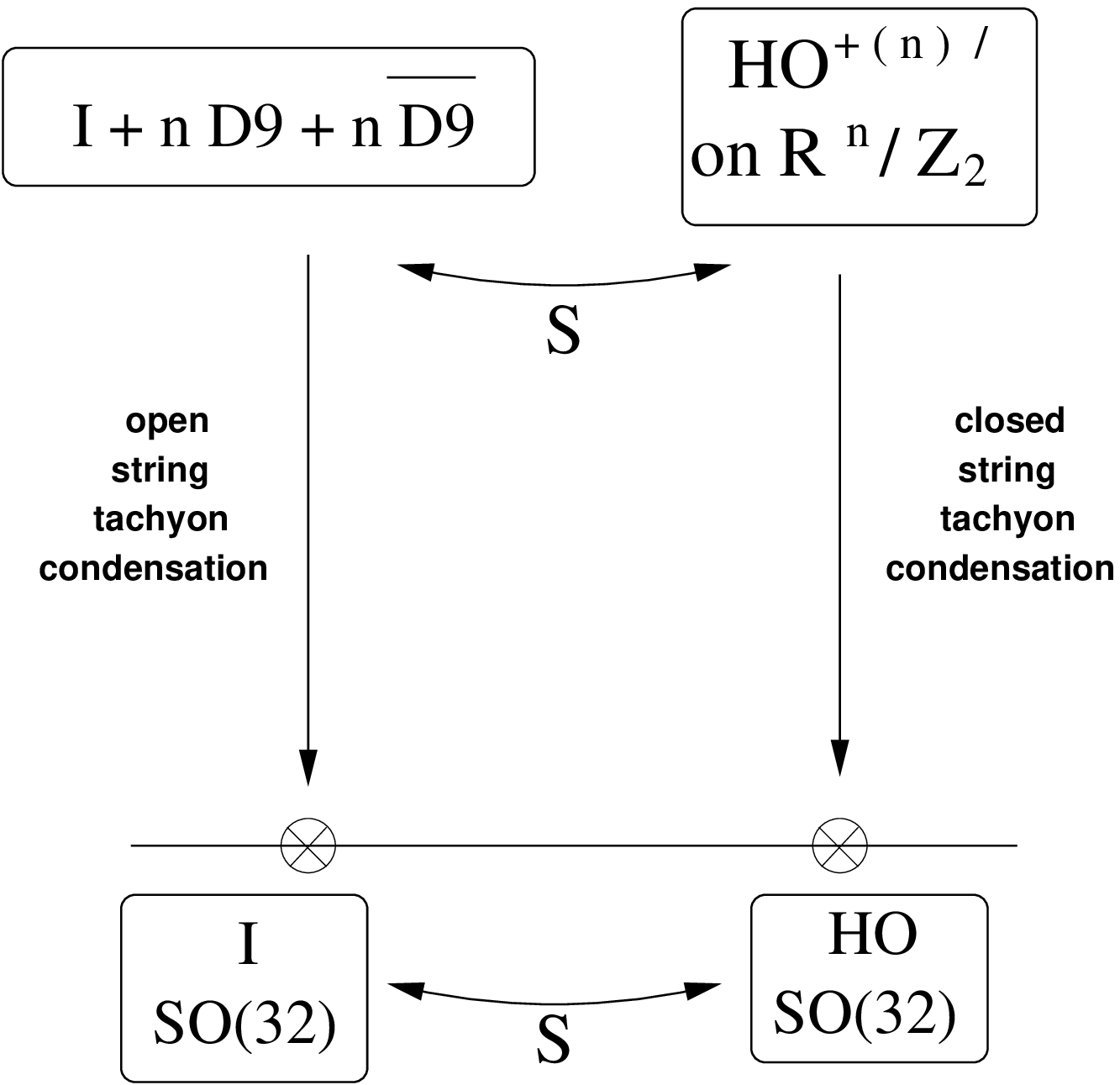}}

If the correspondence is to make sense
as a true S-duality, the higher $SO(n)$ harmonics
of the bulk fields must have some interpretation in
terms of the ten-dimensional open string theory.  The immediate
challenge is that there are single-particle
$SO(n)$ representations such as
$T\uu a\ll{,s\ll 1 s\ll 2 \cdots s\ll {2m+1}}$ which occur
on the $\ho{+(n)}$ side, for which there is no corresponding
$SO(n)$ representation among single-particle states on
the type I side.  The only possible resolution is that 
as the type I coupling is raised, the open string theory
develops a tower of
$SO(n)$ non-singlet bound states, which fill out all the
higher $SO(n)$ representations of spherical harmonics
about the fixed point in the heterotic string theory.
For instance, the heterotic field
\ee{
{{\pp\uu{2m+1} T\uu a}\over{\pp Y\uu{s\ll 1}
\pp Y\uu{s\ll 2}  \cdots \pp Y\uu{s\ll{2m+1}}}}
{}\left . \right \abs
\ll{Y\uu u = 0}
}
has the same $SO(n)$ quantum numbers as the composite
type I field
\ee{
\hat{T}\uu {a \abs s\ll 1 }
\hat{T}\uu{b\ll 1\abs s\ll 2}
\hat{T}\uu{b\ll 1 \abs s\ll 3}
\cdots
\hat{T}\uu{b\ll{m} \abs s\ll{2m}}
\hat{T}\uu{b\ll{m} \abs s\ll{2m+1}}
+\lrd {\rm permutations~of}~s\ll 1,s\ll 2,\cdots,s\ll{2m+1} \rrd
}

So the duality conjecture makes a surprising prediction about the
behavior of the type I theory at strong coupling: it must
enter a \it nonabelian composite phase \rm, in which short-range
forces bind open strings together into new single-particle
states and resonances.
These bound states are very different from the
gauge-singlet 
glueball states which arise from long-range interactions
in confining gauge theories.  In ten dimensions, gauge
interactions are always weak at long distances, so
the binding forces
leave $SO(n)$ color quantum numbers unconfined.

\newsec{Compactification and symmetry breaking}

\subsec{Moduli spaces}

The results of the previous section suggest an S-duality between
type I string theory with $n$ additional D9-$\dnb$ pairs,
and $\ho{+(n)/}$ theory in the presence of a certain kind of
orbifold singularity.  Such a duality would
generalize the well-known
duality between the supersymmetric backgrounds of
the type I and HO string theories.

But we should be careful not to posit an equivalence
between two theories which are manifestly inequivalent;
the type I theory is a ten-dimensional
string theory with ten dimensions' worth of momentum states
for each mode of the string, a finite ten-dimensional
Newton constant and gauge coupling, and so on.

Such a theory, at finite coupling, could not possibly provide
an exact alternate description of a theory living in $10+n$
compact dimensions.

Rather, we propose that at strong coupling, the type I string
with $n$ brane-antibrane pairs
describes the type $\ho{+(n)/}$ string
on a \it family \rm of compact $n$-dimensional spaces,
each of which has a single orbifold singularity of the
type described in the previous section.
The reason that the strong coupling physics of the
type I string is described by a family of type
$\ho{+(n)/}$ theories, rather than a single one,
is that the $\ho{+(n)/}$ theory develops a new branch of
approximate moduli in the limit where the heterotic coupling
is weak: the moduli of the CFT describing the compact
$n$-dimensional space.  If the space is a toroidal orbifold,
for instance, such moduli are exactly massless
at heterotic tree level,
and can be lifted only by string loop effects.

\ifig\phases{
A refined phase diagram illustrating the critical-supercritical
S-duality.  Raising the type I
coupling leads to a potential which breaks
some of the gauge symmetry spontaneously.  The potential
basin describes an approximate moduli space
of toroidal orbifolds with one $\IR\uu n / Z\ll 2$
singularity of stable type.
}
{\epsfxsize4.0in\epsfbox{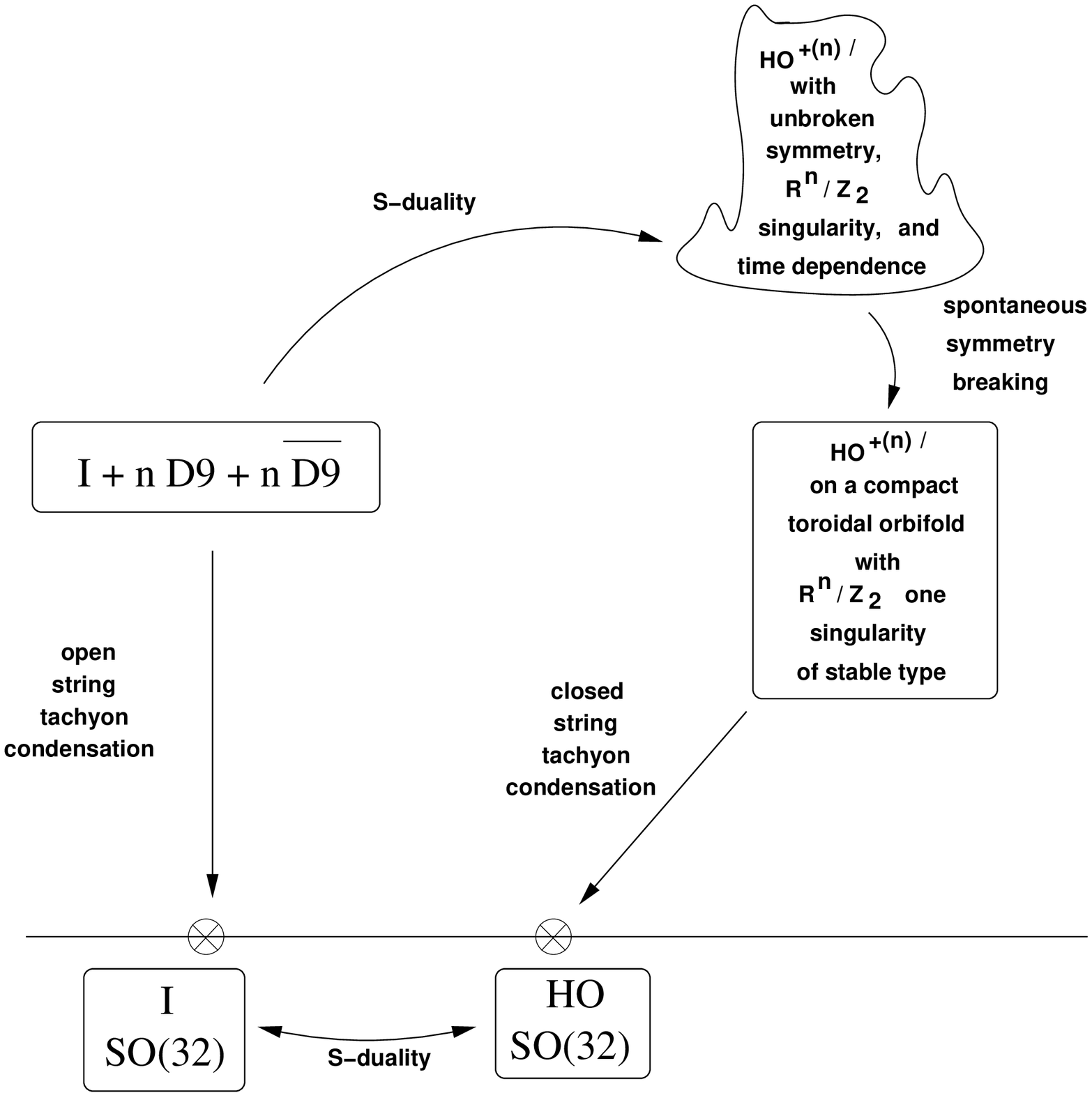}}

In this section we shall construct compactifications
of $\ho{+(n)/}$ theories for $n=1,2$,
 each with a single orbifold singularity
of the stable type we have described in detail.  Each will have
another singularity with different boundary conditions
which break some of the $SO(32+n)$ gauge symmetry spontaneously.
For $n=2$ the compactification itself will break the $SO(n)$
gauge symmetry spontaneously as well.  Nonetheless
in both cases the spectrum of chiral fermions
will be organized into multiplets of the
$SO(n) \times SO(32+n)$ gauge symmetry, and the fermions couple
to the massive gauge bosons according to their
gauge representation.

\subsec{Symmetry-breaking boundaries for $\ho{+(1)/}$}

Let us compactify the type $\ho{+(1)/}$ theory on
an interval.  And we would like to do this in such a way
that the ten-dimensional tachyon potential has a
supersymmetric global minimum describing a single universe in which
the supersymmetric HO theory describes the degrees of
freedom and their dynamics.

There is an obvious way to compactify the theory,
by putting it on an interval with
two boundaries of stable type.  That compactification
cannot describe a system
with a ten-dimensional tachyon which condenses to a
single supersymmetric universe.
In order to have a single universe as the endpoint of
tachyon condensation,
it is important that our compact space have only one orbifold
singularity of stable type.

If we were to have two singularities of stable type, then
upon perturbation by a generic superpotential, the tachyon
will condense to a state with two mutually disconnected
regions. 
To illustrate this, we orbifold a circle $X\uu{10} \sim
X\uu{10} + 2\pi R$ by an inversion $X\uu{10}\to - X\uu{10},
\psi\uu{10}\to-\psi\uu{10}$
combined with $\l\uu a \to - \l\uu a$.  This leaves us with
two boundaries of stable type.  Now perturb this system by
the superpotential
\ee{
W = \l\uu {33} \sin {{X\uu{10}}\over{R}}
} 

The resulting potential $\sin\sqd {{X\uu{10}}\over R}$ has
two zeroes, one at each fixed point.  Each vacuum
has unbroken worldsheet supersymmetry.  The only
massless worldsheet degrees of freedom are $\l\uu{1-32}$
and $X\uu{0 - 9}, \psi\uu{0-9}$, so each vacuum describes
a copy of the worldsheet theory of the supersymmetric
ten-dimensional HO string.

We can also verify that each vacuum contributes with the same
sign to $\tr\lsq (-1)\uu {F_{R_W}}\rsq$ of the effective theory.  To see this,
remember that the right-moving fermion number of the standard
HO theory is the one which inverts $\psi\uu{0-9}$
but does not act on any of the $\l$,
so this symmetry comes from $g\ll 1 g\ll 2$ in the microscopic
theory on the worldsheet.  Under this symmetry $\l\uu{33}$
and $\psi\uu{10}$ are neutral, so despite
the fact that their mass
matrix changes sign as one goes between the two worldsheet
vacua, the two vacua contribute with the same sign
to $(-1)\uu{F_{R_W}}$ in the effective theory.  
The endpoint of tachyon condensation is two stable,
disconnected universes, each with its own graviton,
$SO(32)$ gauge field, and chiral fermions.

To avoid such a situation, we would like to make the second
wall of the interval an \it unstable \rm type of boundary,
with Neumann conditions for at least one of the tachyons.
Let us first consider such a boundary in isolation.

In order to give a tachyon Neumann boundary conditions
at a fixed locus, we need to let the corresponding fermion
$\l\uu a$ be even, rather than odd, under the action that
reverses $X\uu{10}, \psi\uu{10}$.  In order to ensure modular
invariance, we need the number of odd $\l$ to be one plus a
multiple of sixteen.  The simplest way to satisfy this
requirement is to make only one of the $\l\uu a$,
say $\l\uu{33}$, odd at the new wall.  This choice breaks
the $SO(33)$ continuous symmetry down to $SO(32)$.

\subsec{An interval with one boundary of each type}

There are two ways to
study an interval with one boundary of each type.  We could
appeal directly to spacetime locality, and simply
impose the appropriate boundary conditions on the
spacetime fields at each boundary.  This is the approach
taken by the authors of \FlournoyVN.  We could also
obtain the model with one boundary of each type by
starting with the $S^1 / Z_2$ model we described earlier, with
two boundaries of stable type, and further orbifolding by
$g\ll 3$, acting as follows \foot{The conditions for level matching
require that the momentum $p\ll{10}$ be fractional in
winding sectors such as
$g\ll 2 g\ll 3$.  The demonstration of level matching and
closure of the OPE is straightforward but tedious; since winding
strings do not play a role in our discussion, we omit
the details.  Related issues arise for the backgrounds
discussed in \FlournoyVN.} 

\vskip .5in
\bf Table 12: \rm The interval in type $\ho{+(1)}$
string theory with one stable and one
unstable type of boundary can be thought of as
an orbifold by an element $g\ll 3$ of the interval
with two boundaries of stable type.  This table gives
the charge assignments of worldsheet fields under $g\ll 3$. 
\bigskip
\begintable
object | $g\ll 3$ 
\elttt {3 pt} 
$X\uu{10}$ | $X\uu {10} \to \pi R - X\uu{10}$ \elt
$\psi\uu{10}$ |  -  \elt
$\l\uu{1-32}$ | + \elt
$\l\uu{33}$  | - 
\endtable

Such an orbifold describes an interval
of length ${{\pi R}\over 2}$ with
two inequivalent boundaries, one
at $X\uu{10} = 0$ with Dirichlet boundary conditions for
all tachyons, and another at ${{\pi R}\over 2}$ with 
Dirichlet boundary conditions for $T\uu{33}$ and Neumann
boundary conditions for $T\uu{1-32}$.  

Take $R$ to be larger than string scale.
In ten-dimensional terms, the most tachyonic modes have 
$k\uu M k\ll M + 2 i V\ll M k\uu M = -{2\over{\apr}} +
{1\over{4 R\sqd}}$ and transform in the vector representation
of the unbroken $SO(32)$.
They come from acting on the state
$\left \abs \sin {{x\uu{10}}\over {2R}}, k\uu M, 0 \right
\rangle\ll 1$ with
$\l\uu{1-32}\ll{-\hh}$.  
There is a thirty-third scalar
mode which comes from acting on $\left \abs \sin {{x\uu{10}} 
\over R} , k\uu M, 0\right \rangle
\ll 1$ with $\l\uu{33}\ll{-\hh}$.  This scalar satisfies 
$k\uu M k\ll M + 2 i V\ll M k\uu M = -{2\over{\apr}} +
{1\over{ R\sqd}}$ and transforms along with the other
thirty-two tachyons as
the thirty-third component of the vector of the
broken $SO(33)$.

What happens if we condense one of the tachyons which lives
in the $\bf 32$ of the unbroken SO(32)?  The corresponding
relevant superpotential perturbation is 
\ee{
W = \l\uu{1-32} :\sin  {{X\uu{10}} \over {2R}} :
}
For sufficiently large $R$ we can just drop the normal ordering
symbol and treat the perturbation as classical.  The
bosonic potential coming from this perturbation is
\ee{
\sin\sqd {{X\uu{10}}\over{2R}}
}
This potential is nonvanishing everywhere except at the 
left-hand boundary.  In the IR, the theory flows to
a CFT describing strings in a
single universe.

It is easy to see that the string theory in that universe is
the supersymmetric HO theory in ten dimensions.  The massless
degrees of freedom in the infrared are the $X\uu M$ and
$\psi\uu M$, as well as $\l\uu{1-31}$ and $\l\uu{33}$.  

We can also see that it is a stable, supersymmetric universe.
Of the three $Z\ll 2$ factors of the worldsheet gauge group,
$g\ll 3$ is spontaneously broken by the expectation value of
$x\uu {10}$, and $g\ll 1, g\ll 2$ are
unbroken.  On the infrared degrees of freedom
$g\ll 1$ and $g\ll 2$ act
exactly as $(-1)\uu{F_{L_W}}$ and $(-1)\uu{F_{R_W}}$.  So in
the effective theory on the worldsheet there are two independent
$Z_2$ gauge symmetries which act as the chiral $Z_2$ fermion
number symmetry in the supersymmetric, ten dimensional
type HO theory.

If instead of condensing one of the tachyons
in the $\bf 32$ we condense the thirty-third tachyon, 
we get a slightly different result.  The state is
\ee{
\l\ll{-\hh}\uu{33} \left \abs \sin {{x\uu{10}}\over{R}} ,k\uu M,0 \right \rangle
\ll 1
}
and the corresponding superpotential perturbation is
\ee{
\l\uu{33} : \sin{{X\uu{10}}\over R} :
}

This superpotential, and the corresponding bosonic potential,
vanishes in two places, one at each fixed point.

The physics of the universe at $X\uu{10} = 0 $ is still that
of the supersymmetric HO theory, by the same
set of arguments we have already given.  The physics in the
universe on the right is slightly different.

In the universe at $X\uu{10} = {{\pi R}\over 2}$, the unbroken
symmetries are $g\ll 1$ and $g\ll 3$.  $g\ll 1$ acts
as $(-1)\uu{F_W} = (-1)\uu{F_{L_W}} \cdot (-1)\uu{F_{R_W}}$ on
the infrared degrees of freedom, and $g\ll 3$ acts trivially
on all the infrared degrees of freedom.  So the infrared CFT
describes strings propagating in two disconnected universes; in
the first universe the dynamics of
string theory is governed by the critical, supersymmetric
HO, and in the second universe the dynamics are governed by
the critical, nonsupersymmetric $\ho /$.

\subsec{Spontaneous breaking of $SO(n)$: $\ho{+(2)/}$ on
an $\IR\IP\ll 2$ orbifold.}

In the example above, we compactified the theory in a way which
gave rise to a single stable universe after tachyon condensation.
Compactifying the theory in such a way
involved breaking some of the continuous SO(33) gauge symmetry
spontaneously.  We find that this is characteristic
of the duality between unstable brane
configurations and supercritical strings.  The simple
backgrounds we study 
\it cannot \rm be understood simply as limits of the type I
string which one can reach by changing the value of the
dilaton only.
At some point, one must shift the vacuum as well by giving an
expectation value to some field charged under $SO(33)$.  But
if the compactification scale is large, as in the previous example,
the spontaneous breaking can be viewed as small, in the sense that
the masses of the gauge
bosons are small compared to string scale.

Now we shall study an example with $n=2$ and find that
the $SO(2)$ gauge symmetry gets broken as well
as well; to get a background which is static at
tree level (modulo the variation of the
dilaton) and has only ten noncompact dimensions, we must break
some of the $SO(n)$ gauge symmetry spontaneously.  But we will see
that the compactification still has identifiable massive vector
bosons of the Higgsed $SO(2)$ which couple to charged fermions
as a gauge boson should.

We begin with $\ho{+(2)/}$ theory with the dimensions
$X\uu{10,11}$ compactified on a
$T\uu 2$ with gauge bundle $V$.  We define $V$ by letting
$\l\uu{33,34}$ be antiperiodic around both
the $X\uu{12}$ direction
and the $X\uu{11}$ direction.  Such a 
bundle can be obtained as a freely acting orbfold of a torus
of twice the linear size.\foot{
Again, modular invariance requires fractional KK momentum
in winding sectors.  See the previous footnote.
} 
For
simplicity, let the radii $R\ll{10} = R = R\ll{11}$ of
the torus be equal.

\ifig\nonsusyweb{
A space with the topology of $\IR\IP\ll 2$ and a $U(1)$
isometry, with nonzero Ricci curvature and one
$\IR\uu 2 / Z\ll 2$ orbifold singularity of
stable type.  Both the geometric $SO(2)$ and the
current algebra $SO(34)$ are unbroken by this background.  There
are metric and dilaton tadpoles which lead to
spontaneous gauge symmetry breaking.  The pink disc at the top
is a crosscap.  This is the kind of
time-dependent $\ho {+(2)}$ background one might
obtain by taking the type I coupling to be large
in the presence of two D9-branes and two $\overline{\rm D9}$-branes.
We propose that it can relax to a toroidal orbifold background
which is static except for the timelike linear dilaton.
}
{\epsfxsize1.5in\epsfbox{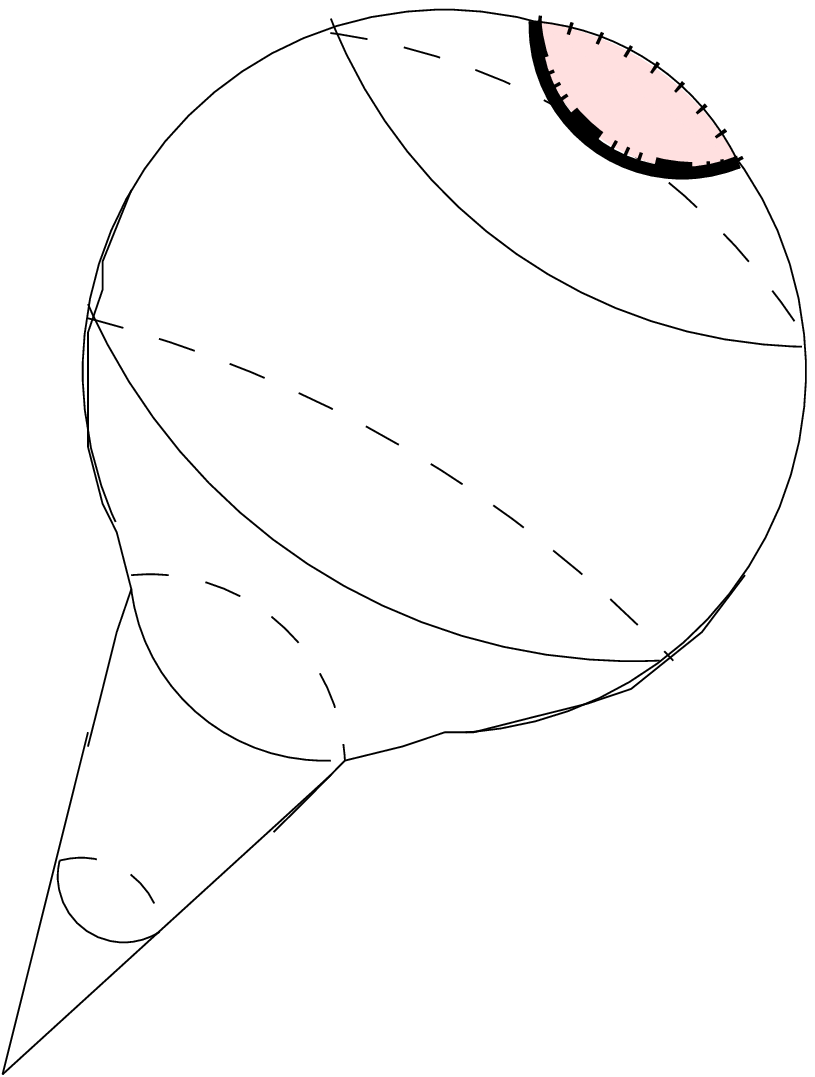}}

Next, we orbifold this space by an action $g\ll 2$ which just
inverts both circle directions simultaneously; $g\ll 2$
also acts with a minus sign on all thirty-four $\l\uu a$.
 This orbifold
has the topology of a sphere and the geometry of a
tetrahedron~\SenVD.
It has four fixed points, two of stable type and two unstable ones.
By 'stable type', we mean what we meant earlier, that all tachyons
at one of the fixed point of stable type $ (X\uu{10}, X\uu{11})
= (0,0 )~{\rm and}~(\pi R, \pi R)   $
have Dirichlet boundary conditions.
At the two unstable fixed points,
\ee{
(X\uu{10}, X\uu{11})
= (0,\pi R)~{\rm and}~(\pi R, 0)
}
the tachyons $T\uu{33,34}$
have Neumann boundary conditions and $T\uu{1-32}$ have Dirichlet
boundary conditions.

Finally we orbifold by another operation $g\ll 3$ which 
exchanges the two fixed points of each type:
\ee{
g\ll 3 : (X\uu{10}, X\uu{11}) \mapsto (X\uu{10} + \pi R~,~
\pi R - X\uu{11})
}
This operation is orientation-reversing and so we let it
act on the fermions as 
\ee{
(\l\uu {33}, \l\uu{34}) \mapsto (\l\uu{33}~,~ - \l\uu{34})
}

Neither $g\ll 3$ nor $g\ll 2 g\ll 3$ has fixed points on the
torus; equivalently, $g\ll 3$ acts freely on the $S\uu 2$.
The action is orientation-reversing and the quotient has the
topology of an $\IR\IP\ll 2$.  Like the interval
described earlier, this orbifold has one fixed point of stable
type, at $(X\uu{10}, X\uu{11}) = (0,0)$.
It has $SO(34)$ and $SO(2)$
unbroken near the fixed point, and Dirichlet
boundary conditions for all the tachyons.

\subsec{Coupling between fermions and gauge field}

When we say that $SO(2)$ is unbroken near the fixed point, this
is more than a manner of speaking.  All string modes can
be organized into $SO(2)$ multiplets
according to the behavior of their
wavefunctions near the fixed point.

First we identify the higgsed $SO(2)$ gauge boson which couples
to the chiral fermions and tachyons according to their
charges of the corresponding fields in the open string
theory.
They are built from the left moving current
\ee{
J \equiv
 J\uu{[10\abs 11]} \equiv
:\sin {{X\uu{10}}\over R} : \pp\ll - X\uu{11}
-  :\sin {{X\uu{11}}\over R} : \pp\ll - X\uu{10}.
}
We will work in the limit where $R$ is much larger than
string scale.  In this limit the breaking of gauge symmetry
is small at the fixed point, and the weight of $J$
is $(1,0) + o({\apr\over {R\sqd}})$.
We can decompose the gauge field vertex operator as
\ee{
\cv\ll{(2)} \equiv \exp{-\phi} ~  c ~\cct~ (e\cdot \psi) ~ 
\exp{i k\ll M\uu{(2)} X\uu M} ~J  .
}

The two fermion vertex operators are
\ee{
\cv\ll{(1,3)} \equiv \exp{-\phi / 2} ~c~\cct~\Theta\ll{\a\ll{(1,3)}}
~\exp{i k\ll M\uu{(1,3)} X\uu M} ~\t\ll{(1,3)},
}
where the $\t$ is the internal piece of the vertex operator,
made out of excited twist fields and current algebra
fermions.  We evaluate the first five factors in the correlator
with ease:
\ee{
\spherecor {\exp{-\phi / 2} }  {\exp{-\phi } }  {\exp{-\phi / 2} }
&= z\ll {12}\uu{- \hh} z\ll{13}\uu{- {1\over 4} } 
z\ll{23}\uu{ - \hh      }
\cr
\spherecor \cct \cct \cct &= \zb\ll {12} \zb\ll {13} \zb\ll {23}
\cr
\spherecor c c c &= z\ll {12} z\ll {13} z\ll {23}
\cr
\spherecor {\Theta\ll{\a\ll{(1)}}} {e\cdot\psi} 
{\Theta\ll{\a\ll{(3)}}} &= {1\over{\sqrt{2}}}
(e\ll\m \G\uu\m \G\uu 0)\ll{\a\ll{(1)} \a\ll{(3)}}
z\ll{12}\uu {-\hh} z\ll{13}\uu{-{3\over 4}} z\ll{23}
\uu{-\hh}  
\cr
\left\langle
\prod\ll{i = 1}\uu 3  \exp{i k\ll M\uu{(i)} X\uu M  } 
\right\rangle\ll{S\uu 2}
 &=  {\cal I}(k\ll {(1)} + k\ll {(2)}
+ k\ll {(3)}) \prod\ll{\matrix {i, j = 1 \cr i < j}} \uu 3 
\abs z\ll{ij}\abs\uu{\apr k\ll {(i)}  \cdot k\ll{(j)} }
}
Here all the formulae are familiar~\PolchinskiRR
~except for the fifth;
the function $\cal I$ represents a formal divergent
integral
\ee{
{\cal I}(k)\equiv
\int d\uu{10} x ~\exp{(i k\ll M - 2 V\ll M )x\uu M}
}
over the zero modes of the embedding coordinates.
To interpret this integral, rewrite $k\ll{(1,2,3)}$ in terms
of the wave vectors of the fields with \it canonical \rm
normalization.  Giving the fields
unit kinetic term involves rescaling them by
$\exp{\Phi}$, which shifts the momentum by $i V$.  So
the wave vectors $k\pr$ of the canonically normalized fields
are related to the wave vectors $k$ of the
fields with conventional string theory normalization by
\ee{
k\ll{(i)} = k\pr\ll{(i)} + i V
}
So the zero mode integral
can be rewritten as
\ee{
{\cal I}\lrd 3 i V + \sum\ll{i = 1}\uu 3 k\pr\ll{(i)} \rrd
= \int d\uu{10} x~ {\rm exp}
\left \{ i \lrd \sum\ll{i = 1}\uu 3 k\pr\ll{(i)}
\rrd \cdot x \right \}  \exp{ V\ll M x\uu M},
}
which is just the standard overlap integral for
wavefunctions of three canonical fields,
with the integrand multiplied by an exponentially growing
coupling constant.

We express the correlator this way in order to
emphasize that the divergence of the amplitude comes
entirely from a zero mode integral, and does nothing more than
encode the enhancement of the interaction vertex by
a coupling which depends on time.

To evaluate the sixth factor of the amplitude, note
that the current $J$ is a primary operator of weight
$ (1,0) + o\lrd {\apr\over{R\sqd}} \rrd$.
Taking $\t\ll{(1)}$ to be the
complex conjugate $\t\ll{(3)}\st$ of $\t\ll{(3)}$,
the easiest way to evaluate the correlator in the
$X\uu{10,11}$ theory is in the operator formalism.

The mode expansion of $J$ in the twisted sector is
\ee{
J\ll j = - {{i\sqrt{\apr}}\over {4R}} \sum\ll{k\in \IZ + \hh }
{{ \at\ll {j - k} \uu{10} \at\uu{11} \ll k 
- \at\ll k \uu{10} \at\uu{11}\ll{j - k}}\over k}
+ o\lrd {\apr\over{R\sqd}} \rrd .
}
The only relevant mode is $J\ll 0$ and the only relevant
terms in $J\ll 0$ are the ones with $k = \pm \hh$.  So we
can write
\ee{
J\ll 0 = -{{i\sqrt{\apr}}\over R} \lrd \at\ll{-\hh} \uu{10} \at\ll\hh\uu{11}
- \at\ll{-\hh}\uu{11}\at\ll\hh\uu{10} \rrd + 
({\rm other~oscillators})+ o\lrd{\apr\over{R\sqd}}\rrd
}
Using the CFT formula
\ee{
&\spherecor {\CO\ll 1} {\CO\ll 2} {\CO\ll 3} =
\cr
c\ll{~\CO\ll 1
,~\CO\ll 2, ~\CO\ll 3}\cdot
z\ll {12} \uu{h\ll 3 - h\ll 1 - h\ll 2}
&z\ll {13}  \uu{h\ll 2 - h\ll 1 - h\ll 3}
z\ll {23}  \uu{h\ll 1 - h\ll 2 - h\ll 3}
\zb\ll {12} \uu{\htt\ll 3 - \htt\ll 1 - \htt\ll 2}
\zb\ll {13}  \uu{\htt\ll 2 - \htt\ll 1 - \htt\ll 3}
\zb\ll {23}  \uu{\htt\ll 1 - \htt\ll 2 - \htt\ll 3}
}
relating correlators on the sphere to structure coefficients of
the OPE,
we can reduce
the three-point function to the calculation of
\ee{
c\ll{~\t\ll{(1)},~ J ,~ \t\ll{(3)}} =
\dbra {\t\ll{(1)}} J\ll 0\ket{\t\ll{(3)}} =
\left\langle \t\ll{(3)}
 \right \abs
J\ll 0 \left\abs \t\ll{(3)} \right \rangle 
}

There are three sets of spacetime fermions, corresponding
to the cases where $\ket{\t\ll{(3)}}$ is equal to
\ee{
\ket {\t\uu{\lrd
\psi\uu{[ab]}\rrd}} &= i ~\l\ll{-\hh}\uu a \l\ll{-\hh}\uu b 
\ket{0}\ll{g\ll 1 g\ll 2}
\cr
\ket {\t\uu{\lrd
\ups\uu{a+ }\rrd}} &= {1\over{\sqrt{2}}}~\l\ll{+\hh}\uu a\lrd 
\at\ll{-\hh}\uu{10} + i \at\ll{-\hh}\uu{11} \rrd
\ket{0}\ll{g\ll 1 g\ll 2}
\cr
\ket {\t\uu{\lrd
\th\uu{++ }\rrd}} &= {1\over{2\sqrt{2}}} \lrd 
\at\ll{-\hh}\uu{10} + i \at\ll{-\hh}\uu{11} \rrd \sqd
\ket{0}\ll{g\ll 1 g\ll 2}
}
All three states are normalized to unity.
A straightforward calculation yields
\ee{
\dbra {\t\uu{\lrd
\psi\uu{[ab]}\rrd}}
J\ll 0
\ket {\t\uu{\lrd
\psi\uu{[cd]}\rrd}} &=  o\lrd{\apr\over{R\sqd}}\rrd
\cr
\dbra {\t\uu{\lrd
\ups\uu{a- }\rrd}}
J\ll 0
\ket {\t\uu{\lrd
\ups\uu{b+ }\rrd}} &= {{\sqrt{\apr}} \over R}
+  o\lrd{\apr\over{R\sqd}}\rrd
\cr
\dbra {\t\uu{\lrd
\th\uu{-- }\rrd}}
J\ll 0
\ket {\t\uu{\lrd
\th\uu{++ }\rrd}} &=  {{2\sqrt{\apr} }\over R} + 
 o\lrd{\apr\over{R\sqd}}\rrd
}
In all three cases,
the leading ${1\over R}$ piece is proportional to the 
$SO(2)$ charge of the fermion.  

These three-point functions encode the tree-level action
for gauge fields and fermions.  So the tree-level effective
fermion
action in ten dimensions
is of the form
\ee{
{i\over{\kappa\sqd\ll{10}}}
\int d\uu {10} x \sqrt{ \abs G\ll{(10)} \abs }
~\exp{-2\Phi\ll{10}}
\lsq \psb \G\uu M \nabla\ll M \psi 
+ \upsb \G\uu M \nabla\ll M \ups
+ \thb \G\uu M \nabla\ll M \th 
+ \Psb\ll N\G\uu M\nabla\ll M \Psi\uu N
 \rsq,
}
where the connection $\nabla$ is covariant with respect
to $SO(2)$ as well as $SO(34)$.  The relationship between
$\Phi\ll {10}$ and the twelve dimensional dilaton $\Phi$
has the usual dependence on the volume of the compactification.
The minimal couplings to the $SO(2)$ gauge field occur with
the same relative coefficients as in the
type I theory with 2 D9+$\overline{\rm D9}$ pairs added~\SchwarzSF.
This consitutes a piece of
evidence in support of our S-duality proposal.

The $O\lrd {\apr\over{R\sqd}}\rrd$ terms represent
corrections to the amplitude due to the spontaneous
breaking of $SO(2)$ symmetry.
Since the $SO(n)$ comes from a geometric rotation of the
supercritical directions, it is unsurprising that
compactification should break this symmetry.  However it is
important to note that this symmetry does organize the fermion
spectrum of the twisted sector into
exactly degenerate multiplets, and that the corresponding
gauge bosons couple appropriately to those fields according to
their $SO(2)$ charges.

\bigskip
\centerline{\bf{Acknowledgements}}
The author would like to thank 
Leonard Susskind, Juan
Maldacena, Jaume Gomis,
Eva Silverstein, 
Michael Gutperle, Brook Williams,
Sergey Cherkis and especially John 
M${^{\underline{\rm c}}}$Greevy for valuable discussions.
We indebted to Michal Fabinger for many
helpful comments on the draft.
I would like to thank the Korean Institute for Advanced Study
for hospitality while this work was in progress.
This work was supported by DOE Grant DE-FG02-90ER40542.

\appendix{A}{Fermion ground states in $\ho +$}

In this appendix we explain how a reality condition
can be imposed on states in the Ramond sectors $g\ll 2$ and
$g\ll 1 g\ll 2$ in a way that is covariant under the
symmetries and consistent with the GSO projection.

We do not do a separate analysis for the bulk fermions
of $\ho {+/}$; all
formulae relevant for bulk fermion states
in that theory can be obtained from the ones
in this appendix by replacing $SO(n)\to SO(n+32)$.
Bulk fermions in the $\ho{+/}$ theory are massive, and
so they come in pairs of the representations discussed here.

\subsec{The $\ho{+(1)}$ theory}

In the Ramond sectors of the
$\ho{+(1)}$ theory the fermion zero
modes satisfy the same Clifford algebra as gamma matrices
with signature $(11,1)$.  The fact that the states
need only be a representation of $SO(10,1)$ and not
the full $SO(11,1)$ allows us to impose a reality condition
on Weyl spinors.

Let $\Gt\uu\m,~\m = 0,\cdots,10$ be a set of gamma matrices
satisfying 
\ee{
\{\Gt\uu\m, \Gt\uu\n\} = 2 \eta\uu{\m\n}.
}
Since eleven is equal to three mod 8 and the signature is
Lorentzian, we can impose the condition that all the
$\Gt\uu\m$ be real.  Then let
\ee{
\G\uu\m \equiv \Gt\uu\m \ot \s\uu 1
}
and
\ee{
\g\uu 1 \equiv 1 \ot \s\uu 3
}
These matrices satisfy the relations
\ee{
\{\G\uu\m, \G\uu\n\} = 2 \eta\uu{\m\n} \llsk\llsk 
\{\G\uu\m, \g\uu 1\} = 0 \llsk\llsk
\{\g\uu 1 , \g\uu 1\} = 2
}
and so together they generate a Clifford algebra of
$SO(11,1)$.
With the identification
\ee{
\psi\ll 0 \uu\m\leftrightarrow {1\over{\sqrt{2}}} \G\uu\m
\llsk & \llsk
\cht\ll 0 \leftrightarrow {1\over{\sqrt{2}}}\g\uu 1,
}
the gamma matrices provide a representation of the canonical
anticommutation relations of the fermion zero modes.

The GSO projection restricts our choice of physical states
to the subspace 
\ee{
64 ~i \lsq  \prod\ll\m \psi\ll 0\uu\m \rsq \cdot \cht\ll 0
\leftrightarrow i \lsq \prod\ll\m \G\uu \m \rsq \g\uu 1 
\equiv \hat{\Gamma}
=\pm 1,
}
where the sign $\pm$ depends on the overall sign of the
GSO projection, as well as on the sign contributed by any
nonzero-frequency oscillators of the $\psi$ and $\chi$
fields with which we act on the ground states.

In the basis we have chosen, 
\ee{
\hat{\Gamma} = 1 \ot \s\uu 2
}
Eigenspinors of this operator cannot be real.  They are
of the form
\ee{
\Psi\ll\pm \equiv
\lrd \matrix { \psi\ll\a \cr \pm i ~\psi\ll\a } \rrd,
}
where $\psi\ll\a$ is a representation of the $\Gt\uu\m$
and with the sign $\pm$ depending on the eigenvalue $\pm$.
The matrix $1\ot \s\uu 1$ permutes the two entries in the
column vector.

Independent of the basis, we know that Weyl
spinors of $SO(8k + 3,1)$ cannot satisfy a covariant
Majorana condition.  However we only need a
Majorana condition which is covariant under $SO(8k+2,1)$,
not the full $SO(8k+3,1)$.  

With that in mind, we impose
\ee{
\Psi\st = \g\uu 1 \Psi.
}
In the basis we have chosen, that condition reduces to the
reality of the spinor $\psi\ll\a$.  
In fact our Majorana condition is actually invariant under
\ee{
\lsq O(10,1) \times O(1) \rsq\ll + = O(10,1) ,
}
which includes reflections across odd numbers of planes,
in addition to
$SO(10,1)$ transformations.  Reflections across a hyperplane
$x\uu \m = 0$ are implemented by the operator
$\G\uu\m \cdot \g\uu 1$, not just $\G\uu\m$.
We use the notation $ \lsq O(10,1) \times O(1) \rsq\ll +  $
to emphasize that the $\cht$ zero mode is also involved
in the orientation-reversing transformations, and also because
it makes the generalization to the higher $\ho{+(n)}$ theories
simpler to see.

The end result for the Ramond sectors of the
$\ho{+(1)}$ theory is that the
states
generated by the action of the fermion zero modes
form a single Majorana spinor of $SO(10,1)$ --
in terms of spacetime field content, 
there are thirty-two real degrees of freedom before equations of
motion are imposed.

\subsec{Generalization to higher $\ho{+(n)}$, with $n\in 2\IZ$}

We break up the problem according to the value of $n$ modulo
eight.

For even $n$, we start with a set of
matrices $\Gt\uu\m$ satisfying the
Dirac algebra
\ee{
\{\Gt\uu\m, \Gt\uu\n\} = 2 \eta\uu{\m\n}.
}
Let 
\ee{
\Gt \equiv i\uu{\lrd 1+  {{(n+1)(n+2)}\over 2} \rrd  }
\Gt\uu 0 \Gt\uu 1 \cdots \Gt\uu{n+9}
}
be the Hermitean chirality matrix of $SO(n+9,1)$.  Similarly,
also define a set $\gt\uu A$ of gamma matrices
of $SO(n)$, satisfying the Dirac algebra
\ee{
\{\gt\uu A, \gt\uu B\} = 2 \d\uu{AB}
}
and define
\ee{
\gt\equiv i\uu{\lrd   {{n(n-1)  }\over 2} \rrd  }
\gt\uu 1 \gt\uu 2 \cdots \gt\uu n
}
to be the Hermitean matrix whose eigenvalue $\pm$ distinguishes
the two
inequivalent spinor representations of $SO(n)$ for even $n$.

From these, we define the matrices
\ee{
\G\uu\m &\equiv \Gt\uu\m \ot 1 \cr
\cr
\g\uu A &\equiv \Gt \ot \gt\uu A \cr
}

With the assignment
\ee{
\psi\ll 0\uu\m \leftrightarrow {1\over{\sqrt{2}}}\G\uu\m \llsk& \llsk 
\cht\uu A\ll 0 \leftrightarrow {1\over{\sqrt{2}}}\g\uu A ,
}
the fermion zero modes satisfy the correct algebra 
\ee{
\{\psi\ll 0\uu\m , \psi\ll 0\uu\n\} = \eta\uu{\m\n} 
\llsk\llsk \{\psi\ll 0\uu\m, \cht\ll 0\uu A\} = 0 \llsk\llsk
\{\cht\ll 0\uu A, \cht\ll 0 \uu B\} = \d\uu{AB}
}

For $n$ even, the GSO projection correlates the
eigenvalues of $\Gt$ and $\gt$ with one another:
\ee{
\Gt\ll{\a\b} \gt\ll{pq} \left \abs  \b,q,({\rm osc.}) \right \rangle\ll{g\ll 2} = 
\pm \left \abs  \a,p,({\rm osc.}) \right \rangle\ll{g\ll 2},
}
where the sign $\pm$ depends on the overall sign choice of the
GSO projection in the Ramond sector, as well as on the
sign contributed by the oscillators.

Next, we consider the reality condition.

Spinors of $SO(n+9,1)$ and of $SO(n)$ have the same reality
properties for all $n$.  For $n \equiv 0$ mod eight, the
Weyl spinors of both $SO(n+9,1)$ and $SO(n)$
are real; for $n\equiv 2$ and $6$ mod
eight, they are complex; and for $n = 4$ they are pseudo-real.

For $n\equiv 0$ (mod 8), we can simply make the matrices $\Gt\uu\m,
\gt\uu A$ all real, and impose the reality condition
\ee{
\Psi\st = \Psi
}
on states.  In this case, the spacetime field content
of a Ramond state (with fixed oscillator content) consists of
two real spinors, say
$\psi\ll{\a p} = \psi\st\ll{\a p}$
and $\psi\ll{\ald, \pd} = \psi\st\ll{\ald,\pd}$, where $\a, \ald$ 
run over bases of the positive and negative chirality
representations of $SO(n+9,1)$ and $p,\pd$ run over
bases of the positive and negative chirality representations
of $SO(n)$.

For $n\equiv 4$ (mod 8), we can do almost the same thing.
Weyl spinors of $SO(n+9,1)$ and $SO(9)$ are
pseudoreal for these values of $n$, which means
there is a conjugation matrix
$B\ll{\a\b}$ (respectively $b\ll{pq}$)
which maps $SO(n+9,1)$ spinors
(respectively $SO(n)$ spinors) of definite chirality
into their complex conjugates
in a way which commutes with the continuous
symmetry group
and satisfies $B\st B = B B\st = - 1$ (respectively
$b\st b = b b\st = -1$).  Further, this map is chirality
preserving, i.e. 
\ee{
\Gt B = B \Gt\st\llsk & \llsk \gt b = b \gt\st
}
Therefore if $\Psi$ is a spinor of the Clifford algebra
generated by $\G\uu\m$ and $\g\uu A$ which satisfies
\ee{
\G\g \Psi = \lrd \Gt \ot\gt\rrd \Psi = \Psi,
}
it is consistent to impose the reality condition
\ee{
\Psi = C \Psi\st,
}
with 
\ee{
C \equiv B\ot b.
}

To understand the meaning of this condition, we choose a basis
in which $\G\equiv \Gt\ot 1 $ and $\g\equiv
1 \ot \gt$ are block diagonal:
\ee{
\G = \lrd \matrix { +1 & & & \cr  & +1 & & \cr & & -1 & \cr
& & & -1 } \rrd \llsk & \llsk \g
= \lrd \matrix { +1 & & & \cr  & -1 & & \cr & & +1 & \cr
& & & -1 } \rrd
}
Writing $\Psi$ in components as
\ee{
\lrd \matrix { \psi\ll{\a p}\cr \psi\ll{\ald
p } \cr \psi\ll{\a \pd} \cr \psi\ll{\ald
\pd } } \rrd ,
}
the GSO constraint $\G \g = \Gt \ot \gt = 1$ says
\ee{
\psi\ll{\ald
p } = \psi\ll{\a \pd} = 0
}
and the reality constraint imposes independent conditions on
the two remaining components:
\ee{
\psi\ll{\a p} = B\ll{\a\b} b\ll{pq} \psi\st \ll{\b q},
\llsk \llsk \llsk \psi\ll{\ald \pd} = B\ll{\ald\bed} b\ll{\pd\qd}
\psi\st\ll{\bed\qd}
}

In terms of spacetime components, this leaves us again with
two spinors $\psi\ll{\a p}$ and $\psi\ll{\ald \pd}$, each of
which satisfies its own reality condition.

For $n\equiv 2,6$ (mod 8) the Weyl spinors of $SO(n+9,1)$ and
$SO(n)$ are neither real nor pseudoreal.  For $SO(n+9,1)$
there is a charge conjugation matrix $B$ with the property
that $\Gt\uu\m B = B \Gt\uu{\m *}$.  By itself this means
that a Dirac spinor of $SO(n+9,1)$ is real or pseudoreal.
But the operation
\ee{
\psi\to B\psi^* 
}
is chirality-changing; if $\psi$ has chirality $\pm$ then
$B \psi^*$ has chirality $\mp$.  This is expressed by the equation
$\Gt B = - B \Gt\st$.  

The $SO(n)$ Clifford algebras for these $n$ also have the property
that there exists a matrix $b$ with $\gt\uu\m b = b \gt\uu{\m *}$,
and $\gt b = - b \gt\st$.

Therefore there is a natural complex-conjutation matrix
for the $SO(n+9,1) \times SO(n)$ cliffod algebra, and it is
\ee{
C = B \ot b 
}
and it has the property that 
\ee{
 (\Gt\ot \gt) C = + C (\Gt\ot\gt)\st
}
and
\ee{
C\st C = C C\st = +1.
}
So it is consistent to impose the condition 
\ee{
\Psi = C \Psi\st 
}
which in terms of components means
\ee{
\psi\ll{\a p} = B\ll{\a\bed} b\ll{p \qd} \psi\st \ll{\bed\qd}.
}
The chirality-changing nature of $B$ and $p$ means that
the $B\ll{\a\b} , B\ll{\ald \bed},
b\ll{pq}, $ and $b\ll{\pd\qd}$ components
of $B$ and $b$ are all zero.  In terms of spacetime field
content, we need never refer to the spinor
$\psi\ll{\ald\pd}$, since we can eliminate it in terms
of $\psi\ll{\a p}$.  So for $n \equiv 2,6$ (mod 8)
we have a single complex Weyl spinor of
$SO(n+9,1)$ which also has definite chirality as a spinor
representation of $SO(n)$.

\subsec{Generalization to $\ho{+(n)}$, with $n\in 2\IZ
+1$}

We construct our gamma matrices as follows.
Let
\ee{
\G\uu\m &\equiv \Gt\uu\m \ot 1 \ot \s\uu 1
\cr
\g\uu A &\equiv 1 \ot \gt\uu A \ot \s\uu 3,
}
where $\Gt\uu\m$ satisfy the Dirac algebra of $SO(n+9,1)$ and
$\gt\uu A$ satisfy the Dirac algebra of $SO(n)$.
The GSO projection restricts us to the subspace
\ee{
1 \ot 1 \ot \s\uu 2 = \pm 1,
}
for some sign $\pm$ depending on the oscillator content
of the state and the overall sign of the GSO projection.
The solutions to the GSO constraint are spinors of
the form
\ee{
\Psi = \lrd \matrix { \psi\ll{\a p} \cr \pm i \psi\ll{\a p}} \rrd
}

For $n\equiv 1,7$ mod eight, the reality condition generalizes
straightforwardly from $n=1$; in these dimensions we can choose
all the gamma matrices to be real and imaginary, respectively.
Then we can impose
\ee{
\Psi = C \Psi\st,
}
with
\ee{
C \equiv 1 \ot 1 \ot \s\uu 3,
}
which is a covariant condition with respect to
$SO(n+9,1)\times SO(n)$.
All pairwise products of $\G, \g$ respect the reality
condition since they are both real
and block diagonal in the last tensor
factor.

For $n\equiv 3,5$ mod eight, the Dirac spinors of 
$SO(n+9,1)$ and $SO(n)$ are pseudoreal, not real.
So there is a conjugation matrix
$B\ll{\a\b}$ (respectively $b\ll{pq}$)
which maps $SO(n+9,1)$ spinors
(respectively $SO(n)$ spinors) into their complex conjugates
in a way which commutes with the symmetry group
and satisfies $B\st B = B B\st = - 1$ (respectively
$b\st b = b b\st = -1$). 

So we find that it is consistent to impose the condition
\ee{
\Psi = C \Psi\st, 
}
with
\ee{
C \equiv B \ot b \ot \s\uu 3.
}

In terms of the spinor $\psi\ll{\a p}$ this condition means that
\ee{
\psi\ll{\a p} = B\ll{\a\b} b\ll{pq} \psi\st\ll{\b q}
}

\subsec{Summary}

We summarize the results of this appendix in a table:

\bigskip
\begintable
n (mod 8) | reality of $\Gt\uu\m,\gt\uu A$ | 
$B\st B$ and
$b\st b$ | $C\ll{SO(9,1) \times SO(n)}$  | field content  
\elttt {3 pt}
0 $~\lrd n\neq 0\rrd$ | real | +1 | $1 \ot 1$  | $\matrix {~~
\cr 
\psi\ll{\a p} = \psi\st\ll{\a p} \cr {\rm and }  \cr
\psi\ll{\ald\pd} = \psi\st\ll{\ald\pd} \cr ~~ } $ 
\elt
1 $~~(n\neq 1)$ | real | +1 | $1\ot 1\ot \s\uu 3$ | $\psi\ll{\a p}
 = \psi\st\ll{\a p}$
\elt
2 |  - | both $\pm 1$ | $B\ot b$ |
$\matrix {~~ \cr
\psi\ll{\a p} 
\cr ~~}
$
\elt
3 | - | -1  
| $B\ot b\ot \s\uu 3$  |  $\matrix {~~ \cr
\psi\ll{\a p} = B\ll{\a\b} ~b\ll{p q}~ \psi\st\ll{\b q} 
\cr ~~}
$
 \elt
4 | - | -1 |
$B\ot b$ | $\matrix {~~
\cr 
\psi\ll{\a p} = B\ll{\a\b}~ b\ll{pq}~
\psi\st\ll{\a p} \cr {\rm and }  \cr
\psi\ll{\ald\pd} = B\ll{\ald\bed} ~ b\ll{\pd\qd}~
\psi\st\ll{\bed\qd} \cr ~~ } $ 
 \elt 
5 | - 
|  -1   | $B\ot b\ot\s\uu 3$  | $\matrix {~~ \cr
\psi\ll{\a p} = B\ll{\a\b} ~b\ll{pq}~ \psi\st\ll{\b q} 
\cr ~~}
$
 \elt
6 | - | both $\pm 1$ |
$B\ot b$ | $\matrix {~~ \cr
\psi\ll{\a p} 
\cr ~~}
$
\elt
7 | imaginary | +1
| $1\ot 1\ot \s\uu 3$ |  $\psi\ll{\a p}
 = \psi\st\ll{\a p}$
\endtable
\bigskip
\noindent

Notes: 

$\bullet{}$
For $n\equiv 2$ (mod 8) the gamma matrices $\Gt\uu\m$
and $\gt\uu A$ can be made real, but this does not
simplify the reality condition; in this basis
$\Gt$ is imaginary and off-diagonal,
and so the $\g\uu A$ are imaginary and off-diagonal
as well.

$\bullet{}$ For even $n$, all spinors $\psi\ll{\a p}$
in the table
should be taken
to have definite chirality with respect to both $SO(n+9,1)$
and $SO(n)$.

$\bullet{}$ The formula for $n\equiv 1$ (mod 8) holds for
$n=1$ only in the sense that we understand the middle
tensor factor to be a $1\times 1$ dimensional matrix,
equal to $1$.

$\bullet{}$ For $n=0$ there is only one chirality
of spinor present, $\psi\ll{\a p} \equiv \psi\ll\a$.  There
is no $\psi\ll{\ald}$.

$\bullet{}$ The symmetry group which acts on spinors
in $\ho{+(n)}$ is a
double cover of $\lsq O(9+n,1)\times O(n) \rsq\ll +$.  The
double cover is an index-two subgroup of a product of
Pin groups: $\lsq Pin(n+9,1)\times Pin(n) \rsq\ll +$,
which is strictly larger than $Spin(9+n,1)\times Spin(n)$.
For more properties of the Pin groups see for example
\BergNE.

$\bullet{}$
The additional elements are generated by reflections across a 
hyperplane $x\uu 1 = 0$ combined with reflection on one
gauge index, say $p=1$.  The action of this element on
spinors is $\Psi \to \G\uu 1 \g\uu 1 \Psi$.
 
$\bullet{}$ For each $n$, the total number of real spinor
components
is $2\uu{n+4}$.
The fermion ground states
form a \it single \rm
irreducible representation of $\lsq Pin(n+9,1)\times
Pin(n)\rsq\ll +$.

\listrefs

\bye